\documentclass[twoside,twocolumn,9pt]{article}
\usepackage{extsizes}
\usepackage[super,sort&compress,comma]{natbib} 
\usepackage[version=3]{mhchem}
\usepackage[left=1.5cm, right=1.5cm, top=1.785cm, bottom=2.0cm]{geometry}
\usepackage{balance}
\usepackage{widetext}
\usepackage{times,mathptmx}
\usepackage{sectsty}
\usepackage{graphicx} 
\usepackage{lastpage}
\usepackage[format=plain,justification=raggedright,singlelinecheck=false,font={stretch=1.125,small,sf},labelfont=bf,labelsep=space]{caption}
\usepackage{float}
\usepackage{fancyhdr}
\usepackage{fnpos}
\usepackage[english]{babel}
\usepackage{array}
\usepackage{droidsans}
\usepackage{charter}
\usepackage[T1]{fontenc}
\usepackage[usenames,dvipsnames]{xcolor}
\usepackage{setspace}
\usepackage[compact]{titlesec}
\usepackage{amsmath}
\usepackage{amssymb}
\usepackage{color} 

\usepackage{epstopdf}

\definecolor{cream}{RGB}{222,217,201}

\usepackage{bm}

\def\be{\begin{equation}}
\def\ee{\end{equation}}
\def\p{\partial}
\def\vp{\varphi}
\def\bn{\textbf{n}}
\def\bi{\textbf{e}}
\def\const{\mathop{\rm const}}
\def\ovm#1{\bgroup \color{blue} OVM: #1\egroup}

\let\Im\imn
\let\theta\vartheta

\begin{document}

\pagestyle{fancy}
\thispagestyle{plain}
\fancypagestyle{plain}{

\fancyhead[C]{
\includegraphics[width=18.5cm]{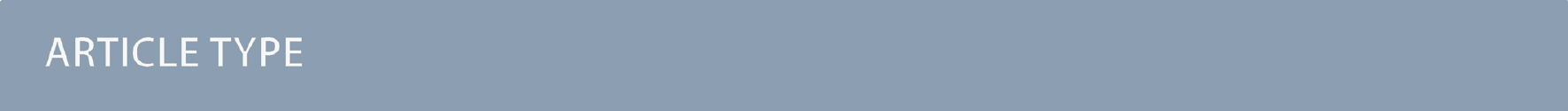}
}
\fancyhead[L]{\hspace{0cm}\vspace{1.5cm}\includegraphics[height=30pt]{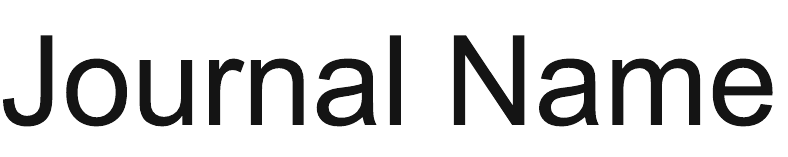}
}
\fancyhead[R]{\hspace{0cm}\vspace{1.7cm}\includegraphics[height=55pt]{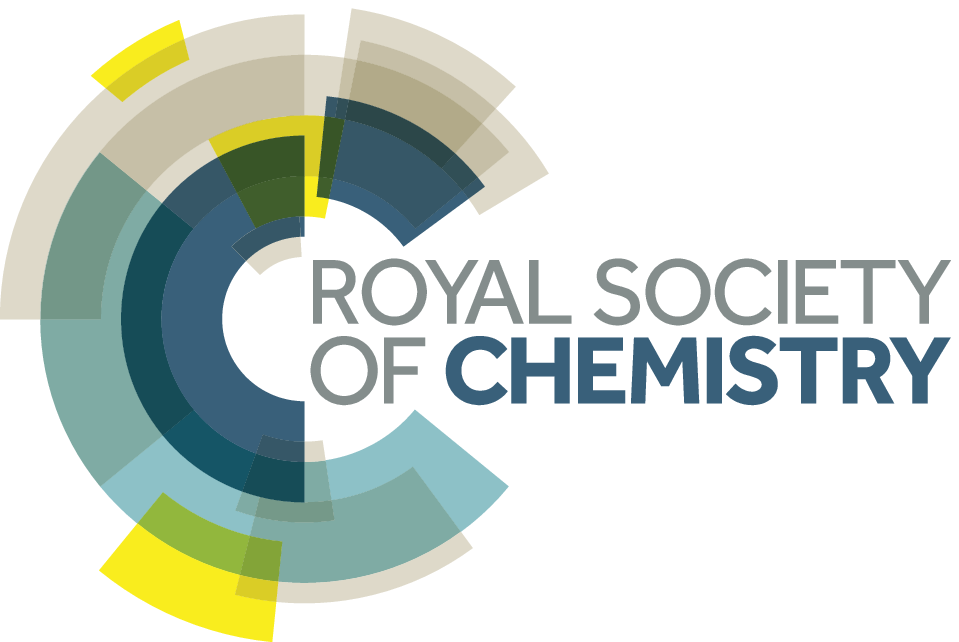}}
\renewcommand{\headrulewidth}{0pt}
}

\makeFNbottom
\makeatletter
\renewcommand\LARGE{\@setfontsize\LARGE{15pt}{17}}
\renewcommand\Large{\@setfontsize\Large{12pt}{14}}
\renewcommand\large{\@setfontsize\large{10pt}{12}}
\renewcommand\footnotesize{\@setfontsize\footnotesize{7pt}{10}}
\makeatother

\renewcommand{\thefootnote}{\fnsymbol{footnote}}
\renewcommand\footnoterule{\vspace*{1pt}%
\color{cream}\hrule width 3.5in height 0.4pt \color{black}\vspace*{5pt}} 
\setcounter{secnumdepth}{5}

\makeatletter 
\renewcommand\@biblabel[1]{#1}            
\renewcommand\@makefntext[1]%
{\noindent\makebox[0pt][r]{\@thefnmark\,}#1}
\makeatother 
\renewcommand{\figurename}{\small{Fig.}~}
\sectionfont{\sffamily\Large}
\subsectionfont{\normalsize}
\subsubsectionfont{\bf}
\setstretch{1.125} 
\setlength{\skip\footins}{0.8cm}
\setlength{\footnotesep}{0.25cm}
\setlength{\jot}{10pt}
\titlespacing*{\section}{0pt}{4pt}{4pt}
\titlespacing*{\subsection}{0pt}{15pt}{1pt}

\fancyfoot{}
\fancyfoot[LO,RE]{\vspace{-7.1pt}\includegraphics[height=9pt]{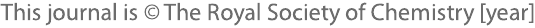}}
\fancyfoot[CO]{\vspace{-7.1pt}\hspace{13.2cm}\includegraphics{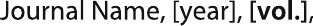}}
\fancyfoot[CE]{\vspace{-7.2pt}\hspace{-14.2cm}\includegraphics{RF}}
\fancyfoot[RO]{\footnotesize{\sffamily{1--\pageref{LastPage} ~\textbar  \hspace{2pt}\thepage}}}
\fancyfoot[LE]{\footnotesize{\sffamily{\thepage~\textbar\hspace{3.45cm} 1--\pageref{LastPage}}}}
\fancyhead{}
\renewcommand{\headrulewidth}{0pt} 
\renewcommand{\footrulewidth}{0pt}
\setlength{\arrayrulewidth}{1pt}
\setlength{\columnsep}{6.5mm}
\setlength\bibsep{1pt}

\makeatletter 
\newlength{\figrulesep} 
\setlength{\figrulesep}{0.5\textfloatsep} 

\newcommand{\topfigrule}{\vspace*{-1pt}%
\noindent{\color{cream}\rule[-\figrulesep]{\columnwidth}{1.5pt}} }

\newcommand{\botfigrule}{\vspace*{-2pt}%
\noindent{\color{cream}\rule[\figrulesep]{\columnwidth}{1.5pt}} }

\newcommand{\dblfigrule}{\vspace*{-1pt}%
\noindent{\color{cream}\rule[-\figrulesep]{\textwidth}{1.5pt}} }

\makeatother

\twocolumn[
  \begin{@twocolumnfalse}
\vspace{3cm}
\sffamily
\begin{tabular}{m{4.5cm} p{13.5cm} }

\includegraphics{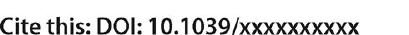} & 
\noindent\LARGE{\textbf{Defect driven shapes in nematic droplets:
analogies with cell division
}
} 
\\
\vspace{0.3cm} & \vspace{0.3cm} \\

 & \noindent\large{Marco Leoni,\textit{$^{a,b,\dag, \ddag}$}  Oksana V. Manyuhina,\textit{$^{a,\dag,\ddag}$} Mark J. Bowick,\textit{$^{a}$} and M. Cristina Marchetti\textit{$^{a}$}} \\

\includegraphics{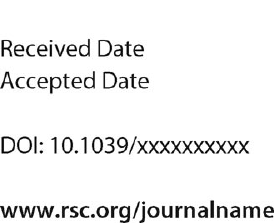} & \noindent\normalsize{Building on the striking similarity between the structure of the spindle during mitosis  in living cells
and nematic textures in confined liquid crystals,  
we use a continuum model of 
two-dimensional nematic liquid crystal droplets, 
to examine the physical aspects of  cell division.
The model investigates the interplay between bulk elasticity of the microtubule assembly, described as a nematic liquid crystal, and surface elasticity of the cell cortex, modeled as a bounding flexible membrane, in controlling cell shape and division. The centrosomes at the spindle poles correspond to the cores of the topological defects required to accommodate nematic order in a closed geometry. We map out the progression of both healthy bipolar and faulty multi-polar division as a function of an effective parameter that incorporates active processes and controls centrosome separation.  A robust prediction, independent of energetic considerations, is that the transition from a single cell to daughters cells occurs at critical value of this parameter. Our model additionally 
suggests 
that microtubule anchoring at the cell cortex 
 may play an important role 
for successful bipolar division. This 
can be tested experimentally by regulating microtubule anchoring.
} 

\end{tabular}

 \end{@twocolumnfalse} \vspace{0.6cm}

ates  ]

\renewcommand*\rmdefault{bch}\normalfont\upshape
\rmfamily
\section*{}
\vspace{-1cm}


\footnotetext{
\textit{$^{a}$~
Physics Department, Syracuse University, Syracuse, NY 13244, USA.
 }
 }
\footnotetext{
\textit{$^{b}$~Institut Curie, PSL Research University, CNRS,
UMR 168, 26 rue d'Ulm, F-75005, Paris, France. }
}
\footnotetext{\dag~`Present address:' mleoni@syr.edu; omanyuhi@syr.edu}
\footnotetext{\ddag~
`These authors contributed equally to this work' }


\section{Introduction}
Living cells have the remarkable  ability to divide, generating  two daughters from a single mother cell. In eukaryotes this  process is termed
mitosis~\cite{Albert}.
A proper division process  provides descendants with the correct amount of genetic material needed for subsequent cell growth and development, thus ensuring the continuation of the life cycle. Failure in the division process can lead to  malfunctioning daughter cells that may  give rise to cancer in multicellular organisms~\cite{Brinkley,godinho}. 
A key role in mitosis is played by the spindle, a self-organized subcellular structure composed of microtubules that segregates chromosomes during the division process and has been previously described as an active nematic gel. 
In normal conditions,  upon entering mitosis cells possess a single pair of centrosomes~\cite{stearns:2015}  that serve as  microtubule nucleating and organizing centers to build the spindle and are driven to opposite sides of the cell during the division process~\cite{Albert}. 
There are some circumstances, however, in which cells may accumulate more than two centrosomes~\cite{Brinkley,godinho,Marthiens,Basto-cell,Duncan}, perhaps  as a result of genetic mutations or previous failed divisions. The precise number and location of centrosomes determines the organization of the spindle, which can be either bipolar or multipolar. 
 Extra centrosomes have been shown to initiate tumorigenesis in flies~\cite{Basto-cell} and result in reduced brain development (microcephaly) in mice~\cite{Marthiens13}.

\begin{figure}[t]
\centering
\includegraphics[width=.7\linewidth]{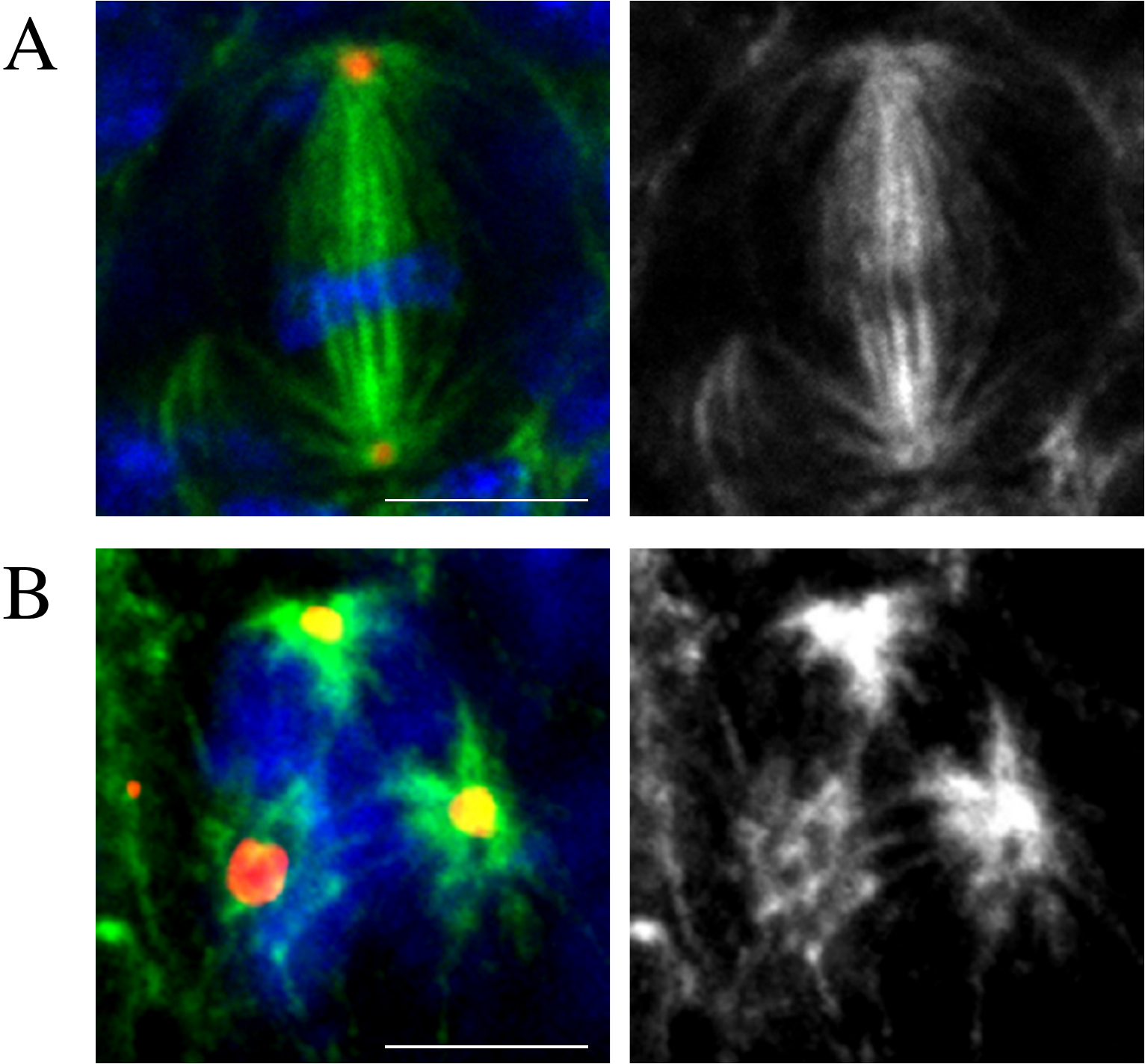}
\caption{\label{fig:spindle}
Images of {\it Drosophila} third instar larval neural stem cells (neuroblast), labelled with antibodies against tubulin (shown in green) and centrosomes (shown in red). DNA is shown in blue. 
A: Wild-type neuroblast in metaphase showing a bipolar microtubule array emanating from both centrosomes. B: A mutant neuroblast that contains extra centrosomes forms a tripolar spindle. Image courtesy of Maddalena Nano from the Basto lab, at UMR144 Institut Curie, Paris, France. 
Scale bar $= 7~\mu$m. 
}
\end{figure}

To change shape and generate forces during the various phases of division a cell has to coordinate multiple signals, of both mechanical and chemical origin, on broad time and length scales. 
To accomplish this task, the cell 
undergoes a local re-organization of microtubules, which  arises spontaneously during the spindle assembly, 
or can 
be induced by external constraints, such as micropatterned environments~\cite{Thery}. An assembled spindle  
usually 
consists of two poles with microtubule stemming radially from the poles forming two asters (see Fig.~\ref{fig:spindle}A).  
Microtubules extending outwards form the spindle  are normal to the cell cortex,  while in the middle of the spindle the microtubules are  aligned tangentially to the cell membrane. The organization, shape, and dynamics  of a bipolar spindle  are fairly well  captured in the framework of active nematic liquid crystals~\cite{Brugues}, which in the presence of topological defects exhibit spontaneously generated flows induced by active stresses~\cite{Sanchez2012,Giomi2013} that can drive droplet division~\cite{Giomi}. 
In the final stages of the division process, a key role is also played by the actin cortex, which lies underneath the cell membrane and organizes the acto-myosin contractile ring that pinches apart the two daughter cells~\cite{Salbreux,Turlier,Paluch}{~\cite{mikko}}. The formation of the contractile ring has also been modeled as arising from the
spontaneous organization of an active nematic gel~\cite{Salbreux,Turlier}. 
\looseness=-1

Inspired by this previous work, we liken the spindle of a dividing cell to a  droplet of nematic liquid crystal bounded by a flexible membrane representing the cell cortex. 
The structure of the  microtubules in the spindle is described by the orientational order of the nematic, with the spindle poles  corresponding to the defect structures that are required by topology when confining a nematic droplet. We then model microtubule organization and cell shape changes during division by examining the interplay of bulk and boundary elasticity in controlling the energetics of defect textures as a function of the separation of the spindle poles. Using entirely analytical methods, we show that the nematic droplet evolves from a singly connected structure of evolving shape (a single cell) to two disconnected structures (the two daughter cells) as a function of a single dimensionless parameter that controls spindle pole separation and the associated cell shape. \looseness=-1

Of course cell division is  an out-of-equilibrium process, with spontaneous organization and shape changes that occur over  time and are driven by complex active processes inside the cell.  Our model does not explicitly describe nonequilibrium active processes. Instead the active processes that drive cell division are incorporated via a single effective parameter $\xi$ proportional to the spindle poles separation during bipolar division.  We assume that the time scales associated with the active rearrangements in the spindle structure are fast compared to those controlling the time evolution of the pole separation, and treat the spindle structure itself as a passive nematic, implicitly incorporating all active processes in the time dependence of $\xi(t)$. 
 Using this minimal effective model, we then evaluate the energy of our ``dividing droplet''  as a function of cell shape, as tuned by $\xi$, and controlled by the interplay of the bulk nematic elasticity of the spindle,
the bending rigidity and tension of the boundary that models the cell cortex, and the strength of the anchoring of the microtubule nematic to the cell cortex. 
We show that a critical value of $\xi$ separates singly connected from divided droplets. Importantly, the prediction of this  critical value is robust, independent from energetic considerations.  Additionally, we propose an energetic criterion for successful division, and use it to construct a phase diagram of cell shapes during the division process.

 Our model is two-dimensional and accounts for the interplay of bulk and cortex elasticity. It can describe both healthy (bipolar) cell division, as well faulty divisions driven by multipolar spindles.
Within our model anchoring of 
microtubules
at the 
 droplet
boundary is necessary for successful bipolar division, 
suggesting that a similar scenario may hold for living cells. 
This prediction could be tested experimentally for example by regulating the expression of proteins -- e.g., of the EB family~\cite{EBprotein} -- that control the attachment of microtubules to the cell cortex. It also suggests that the coupling of microtubules to the cell cortex must be incorporated for understanding the mechanics of the division process.

By examining tripolar spindles, we quantify the conditions that control whether the cell divides into three identical daughter cells or into two cells in terms of an angle characterizing the geometry of the triangular arrangements of the three centrosomes. 
 Remarkably, our prediction for this angle is consistent with experimental observations in populations of \textit{Drosophila}~cells~\cite{Sabino}.

\section{Model}

 Inspired by images of cell mitosis like the ones shown in Fig.~\ref{fig:spindle}, we explore the analogy between the division of biological cells and the interplay of elasticity and surface anchoring in controlling the shape of nematic liquid crystal droplets.  Focusing on the two-dimensional cross-section of a  cell, the star-like assembly of microtubules around the  nucleus before mitosis resembles the $+1$  defect that is required by topology in a circular nematic droplet with normal anchoring. When the cell starts dividing the microtubule arrangement in the bipolar spindle resembles two $+1$ defects separated by a growing distance, while the cell boundary deforms. To ensure conservation of 
 the
 topological charge a total $-1$ charge is then distributed along the cell boundary through variations in the orientation of nematic order relative to the spatially varying normal~\cite{SM:2015}.  To quantify the analogy between dividing cells and liquid crystal droplets we map the microtubule assembly onto the nematic director field and the centrosomes that drive the spindle organization (displayed as localized red regions in Fig.~\ref{fig:spindle}) onto the cores of topological defects in the liquid crystal. We then examine how nematic elasticity  competes with the surface anchoring of the microtubules at the deformable cell membrane and the required defects  in controlling cell shape and topology. The paradigm proposed below is capable of accounting for both healthy (bipolar) mitosis as displayed in the upper row of Fig.~\ref{fig:spindle}, as well as abnormal mitosis, when the spindle can divide in three or more parts, as shown in the bottom row of images of Fig.~\ref{fig:spindle}. 
In the case of three or more centrosomes the splitting of the nuclear content  
  is generally accompanied with 
  an
abnormal 
  DNA 
  configuration
  in a region void of microtubules (the blue areas in Fig.~\ref{fig:spindle}). In our model these regions correspond the negative-charge defects, for instance a $-1/2$ disclination at the center of a tripolar spindle. \looseness=-1

We restrict ourselves to two-dimensional droplets that could describe
cells of height  much smaller  than their lateral extent, as is for instance the case for
epithelial cells.
We ignore  (de)polymerization dynamics of microtubules, which occurs on timescales of the order of seconds, and model cell shape changes,
which can occur on timescales of hours  and are associated with
our parameter $\xi$, as an adiabatic process~\cite{Albert,MTdyn} 
 as outlined in the introduction. 
 
The free energy of our nematic droplet (a.k.a. model cell) is given by
\be\label{eq:ftot}
F\!=\!K\!\!\int_\Sigma\!\! d^2x \underbrace{|\nabla \bn|^2}_{\scriptsize\mbox{elastic}}+\!\!\int_{\p\Sigma}\!\!\!\!ds\,[\underbrace{k_b(\nabla\cdot\pmb\nu)^2}_{\scriptsize\mbox{bending}} +\!\!\!\underbrace{\!\!\gamma_c\!\!}_{\scriptsize\mbox{tension}}\!\!\!-\underbrace{W_a(\bn\cdot\pmb\nu)^2}_{\scriptsize\mbox{anchoring}}].
\ee
The first term on the right hand side of \eqref{eq:ftot} describes the  cost of elastic deformation of a $2D$ nematic droplet spanning a domain $\Sigma$, with isotropic elastic constant  $K$ and unit vector  $\bn$ characterizing the local orientation of microtubules~\cite{Degennes}. The droplet is bound by a flexible boundary $\partial\Sigma$ of bending rigidity $k_b$ and line tension $\gamma_c$ describing the cell cortex. Finally, the last term incorporates the strength $W_a>0$ of the anchoring of the nematic to the boundary, with $\pmb\nu$ the unit normal pointing towards the outside of the cell.

Assuming a cortex thickness of $\sim 100$~nm, the membrane properties  in~\eqref{eq:ftot} can be estimated from experimental measurements  as $k_b \propto 10^{-19} \div 10^{-16}$~J; $\gamma_c \propto 5 \cdot 10^{-3}$~J/m$^2$~\cite{sens:2015}.
There is, however, no direct information available for $K$ and $W_a$.  Typical values for liquid crystals are 
 $K\simeq 10^{-11}$~J/m and~$W_a \propto10^{-7}-10^{-3}$~J/m$^2$~\cite{Degennes}, but these values are likely to be quite different in living cells. Below we show how our work can be used to infer a prediction for the value of $W_a$.

\begin{figure}[t]
\centering
\includegraphics[width=\linewidth]{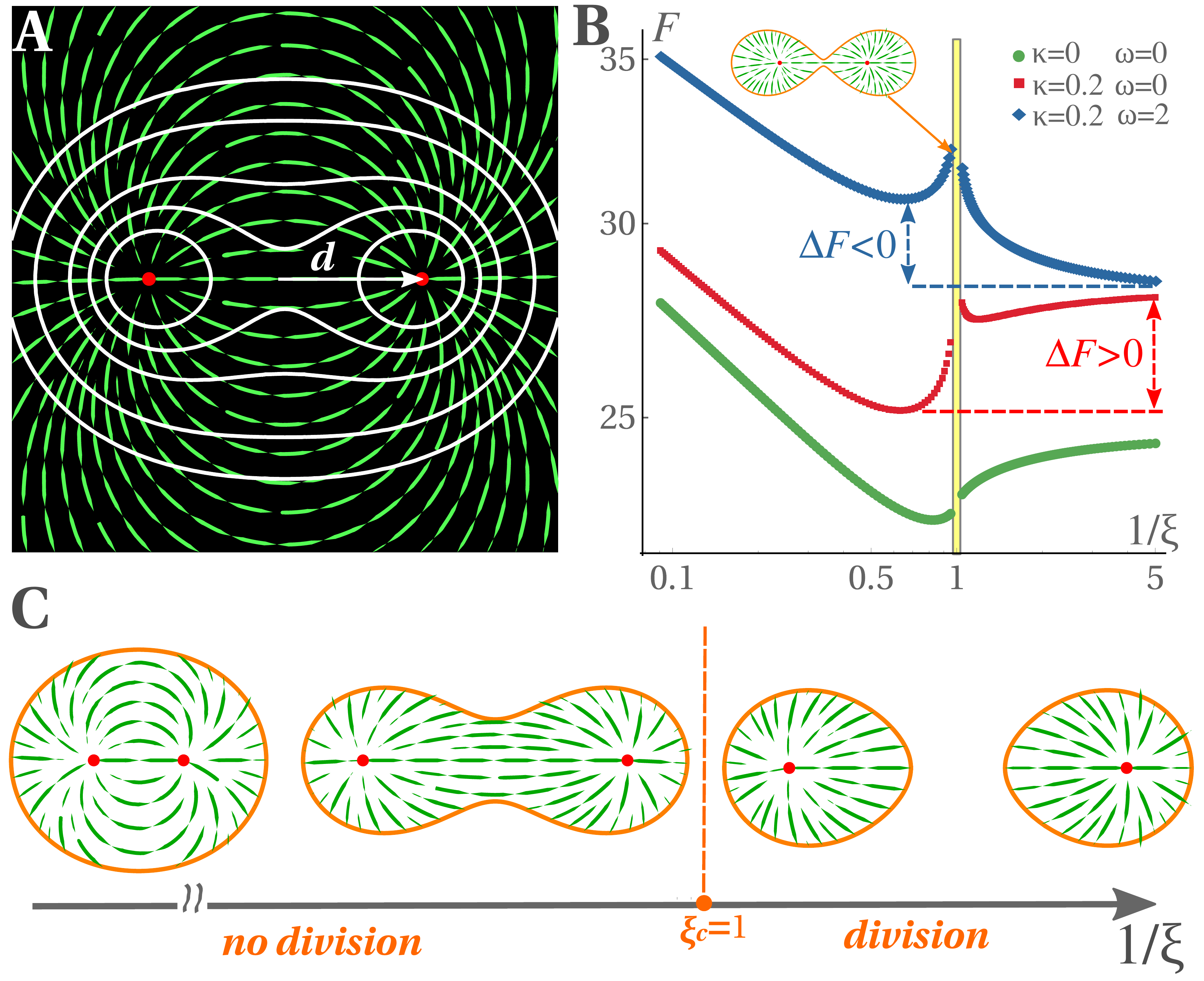}
\caption{\label{fig:field2} A. Nematic structure corresponding to two $+1$ defects separated by a distance $d$. The green dashed lines  describe the direction of the nematic director~$\bn$, representing orientation of the microtubule  in a bipolar mitotic spindle. The red dots are the cores of the topological defects  representing the centrosomes. The solid white lines are the family of boundaries that minimize the energy with stress free boundary conditions (\eqref{eq:level2}), with constant $d$. B.~The free energy~\eqref{eq:ftot} as function of the shape parameter $\xi^{-1}$, for the the defect configuration shown in A,  with defect core size  $\epsilon=0.1R_0$ for various values of $\kappa$ and $\omega$, with $\gamma=0$. We define $\Delta F  = F_f- F_m$, where  $F_m$ is the minimum free energy before division ($\xi^{-1} <1$) and $F_f$ is the final energy for $\xi^{-1} \gg 1$. The yellow region excludes the singularity at the critical value $\xi_c=1$ associated with the pinching-off of cells at division and accounts for the finite thickness of the cortex not incorporated in our continuum model.  C.~Shapes of nematic droplets with constant area: circular ($\xi\gg1$); peanut ($\xi\simeq1.2$ minimum of the green curve in~B); divided ($\xi<1$).}
\end{figure}

Our goal is to determine the shape of the nematic droplet as a function of the configuration and topology of the enclosed defects  by minimizing the free energy,~\eqref{eq:ftot},
with respect to the angle $\alpha$, with  $\bn=\cos\alpha\bi_x+\sin\alpha\bi_y$. 
The resulting equations~\eqref{eq:bc},  given in the Appendix~\ref{sec:Appendix-boundary-condition},  cannot, however, be solved analytically. To proceed 
we first neglect anchoring and simply minimize the elastic (bulk) part of the nematic free energy with  stress-free boundary conditions, which gives
\be\label{eq:stressfree}
\nabla^2\alpha=0, \quad \,\left[{\pmb\nu} \cdot \bm\nabla \alpha\right]_{\partial\Sigma} =0\;,
\ee
Equations~(\ref{eq:stressfree}) yield a family of solutions for possible cell shapes consistent with a prescribed defect configuration. We will then assume that the physical solution can be identified with the one from this family that minimizes the full free energy,~\eqref{eq:ftot}, including  the anchoring energy.  The validity of this approximate method is tested via a perturbative scheme in the Appendix~\ref{sec:Appendix-perturbative}.

It is convenient to rewrite \eqref{eq:stressfree} in the complex plane as 
\be\label{eq:complex}
\p_z\p_{\bar z}\alpha=0, \quad \,\left[\partial_z\alpha e^{i\vp}+\partial_{\bar z}\alpha e^{-i\vp}\right]_{\partial\Sigma}=0\;,
\ee
with $z = x + i y$, $\bar z=x-iy$ and $\vp$ the angle of  $\pmb \nu$ with the $x$-axis. For a configuration of the director field 
with defects of strength   $q_k$ located at positions $z_k$, the angle $\alpha$ can then be written in terms of a potential $\Omega(z)=\Pi_k(z-z_k)^{q_k}$ as 
\be\label{eq:bulk}
\alpha=\left[\log\Omega(z)-\log\Omega(\bar z)\right]/(2i)\;.  
\ee
For instance, for a single $+1$ 
defect at the origin we have $\Omega=z$. In this case the stress-free boundary condition, given by the second of \eqref{eq:stressfree},  is satisfied by a set  of concentric circles of radius $R_0$ in the complex plane defined by $z\bar z = R^2_0$.
More generally,  the boundary condition requires $\mathcal{F}(\{z_k\})\equiv\left[\Omega(z)\cdot\Omega(\bar z)\right]_{\partial\Sigma}={\rm constant}$ (see Appendix~\ref{sec:Appendix-boundary-solution} for details) and determines the shape of the boundary $\p\Sigma$ for a  given defect configuration $\Omega(z)$. For  symmetric defect configurations,  consisting  of positive $+1$ defects at equal distance $d$ from the center of the structure, like the bipolar, tripolar and quadrupolar arrangements shown in Figs.~\ref{fig:field2}, \ref{fig:field3}, \ref{fig:field4}, the shape of the boundary $\p\Sigma$ is determined by the boundary condition that can be written in the form~\footnote{For the tripolar configuration the left hand side of \eqref{eq:xi} additionally depends on an angle that characterizes the shape of the triangle.}
\be\label{eq:xi}
\left[\Omega(z/d)\cdot\Omega(\bar z/d)\right]_{\partial\Sigma}=\xi\;,
\ee
where
$\xi$ is a dimensionless parameter that determines the family of allowed shapes.

In the next section we employ this method to describe the family shapes obtained from \eqref{eq:bulk} and \eqref{eq:xi} for the case of two $+1$ defects corresponding to healthy bipolar division and demonstrate the role of anchoring energy in shape selection.

\subsection{Shapes with two defects: bipolar division}

Here we apply the method to describe shape selection for a nematic droplet with a symmetric arrangement of two $+1$ defects separated by a distance $2d$, as shown in Fig~\ref{fig:field2}, with $\Omega(z)=(z-d)(z+d)$. 
Using polar coordinates    $(r,\theta)$,
with $z= r e^{i \theta}$,  the family of allowed boundaries obtained from \eqref{eq:xi} with
stress free boundary conditions is given by
\be
\label{eq:level2}
 r_{\pm}(\theta)= d\sqrt{\cos(2\theta)\pm\sqrt{\xi -\sin^2(2\theta)}}\;.
\ee
The 
corresponding curves are shown in  Fig.~\ref{fig:field2}A  for a range of values of $\xi$.  
When $\xi>1$ the droplet is a simply connected domain defined by $r_+$, while for $\xi<1$ the droplet
splits into two smaller domains, outlined by the radius vectors $r_{\pm}$ 
(see Fig.~\ref{fig:field2}C).

In order to select from this family of equipotential boundaries those that correspond to possible  shapes of a cell with two centrosomes we now proceed as follows. First, instead of looking at shapes of varying area for fixed value of the defect separation $d$, we impose cell incompressibility in $2D$, i.e., require the cell area to be constant and allow the distance $d$ to vary, as shown 
 in  Fig.~\ref{fig:field2}C\,\footnote{Cells are generally treated as incompressible in $3D$.  Our model is strictly two-dimensional and represents 
  a $2D$ analogue where the cell is treated as a $2D$ object and 
 height fluctuations are neglected.}.
 This amounts to considering $d$ itself to be an increasing function of $\xi$ (see for details Appendix~\ref{sec:Appendix-incompressibility},~\eqref{eq:s1} and~\eqref{eq:s2}), with the shapes becoming more elongated with increasing  $d$. This amounts to assuming that the parameter $\xi$ incorporates all active processes that drive cell division and sets the centrosome separation $2d(\xi)$. Secondly, we incorporate boundary tension and bending as well as anchoring by calculating the full free energy as given in \eqref{eq:ftot} for a given cell shape and using the behavior of the energy to define a criterion for identifying preferred cell shapes.

 Measuring lengths in units of the typical size $R_0 \sim 10~\mu$m of the cell  and energies in units of the nematic stiffness $K$, the free energy depends on three dimensionless parameters: the scaled bending rigidity $\kappa = k_b/(K R_0)$ and tension $\gamma =\gamma_c R_0/K$ of the cell boundary, and 
 $\omega = W_a R_0/K $ that measures the relative strength of anchoring and nematic elasticity.   The full free energy given in~\eqref{eq:ftot} corresponding to a cell configuration~$d(\xi)$ is shown in Fig~\ref{fig:field2}B as a function of $\xi$ for a few parameter values. For  $\xi^{-1} \ll 1$ the cell is almost round, just before entering the anaphase~\cite{Albert}. The free energy has a local minimum $F_m$ for $\xi^{-1}<1$ and it diverges at $\xi=1$. This divergence is nonphysical and is cutoff by the finite thickness of the 
droplet boundary
(cell membrane)
 that has been neglected in our model. This is indicated by the yellow vertical boundaries in Fig.~\ref{fig:field2}B.  
 Based on the analogy between cells and nematic droplets outlined above, we
 conjecture that local minimum of the free energy for $\xi^{-1} < 1$  corresponds
to the most stable cell shapes at the end of anaphase where the two centrosomes are already well separated, and that  the energy barrier 
at $\xi=\xi_c = 1$  represents the energy needed to break the quasi-pinched droplet into two pieces.
In living cells this energy can be provided at the onset of cytokinesis by the cytokinetic ring, which effectively acts as a rope  pulled by active forces, generating a constriction at the cell midpoint~\cite{Turlier,Salbreux}. 
We further conjecture that 
a successful  division is governed by the sign of $\Delta F=F_f-F_m<0$, where $F_f=F(1/\xi\rightarrow\infty)$ is the final free energy for $\xi^{-1}\gg1$, and that
cell division will occur only 
if $\Delta F<0$. This condition is satisfied, for instance,  for the parameter values corresponding to the 
blue curve of Fig.~\ref{fig:field2}B, but not for the other two, where $\Delta F>0$. \looseness=-1

\begin{figure}[t]
\centering
\includegraphics[width=\linewidth]{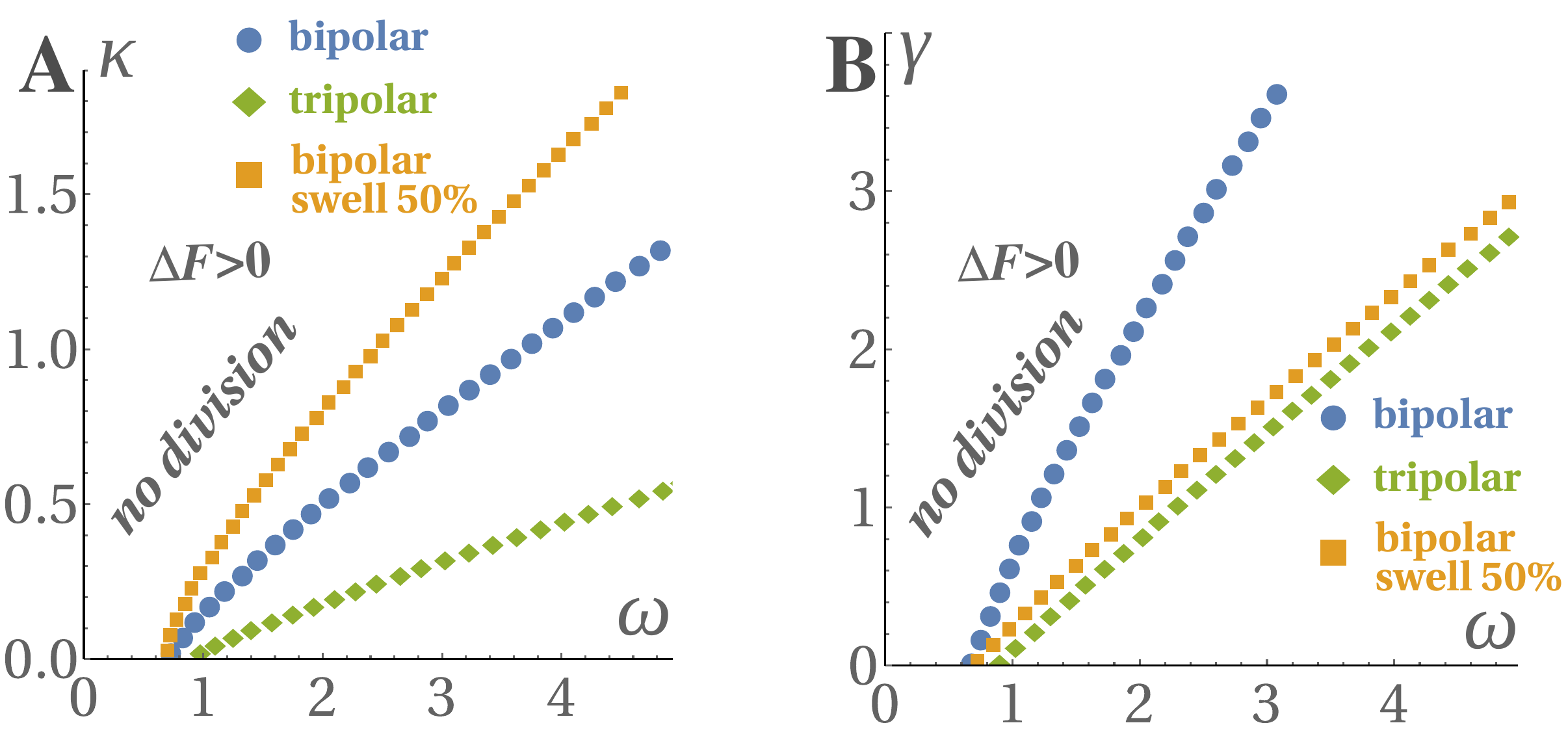}
\caption{\label{fig:phase-diagram}
Phase diagrams for a nematic droplet with bipolar (blue dots) 
and symmetric tripolar (green diamonds,  with  $\beta=\pi/3$, see next section)
division  assuming constant surface area enclosed by the droplet for $\gamma=0$ (A) and $\kappa=0$ (B).  Droplet division (according to our criterion, $\Delta F<0$) occurs in the region 
 between 
the curves and the $\omega$ axis and requires a finite  value of anchoring $\omega \neq 0$ at the cell boundary.  The orange squares represent the division boundary obtained when
swelling of both daughter droplets  and centrosome area by $50\,\%$ is included.}
\end{figure}
Using this criterion, we have constructed phase diagrams displayed in Fig.~\ref{fig:phase-diagram} identifying the regions of parameters  corresponding to successful cell division.  
The  region where division occurs is enclosed between the line and the $\omega$-axis. Remarkably, we find that in all cases, 
anchoring  of filaments to the cell boundary,  $\omega \sim O(1)$, is necessary for division.

Using experimental values of $k_b$ and $\gamma_c$, our model can be used to infer the value of the anchoring strength $W_a$ in two ways.
First, when line tension is negligible ($\gamma\ll1$)
 cell division occurs for values of $\omega$ such that   $\omega \sim \kappa$ (blue dots in Fig.~\ref{fig:phase-diagram}A), corresponding to $W_a \sim k_b/R^2_0 $. For   $R_0\sim10~\mu$m,     
we would then estimate  $W_a \sim 10^{-9}\div 10^{-6}$~J/m$^2$.  Secondly, in the opposite limit $\gamma\gg \kappa$ (expected e.g. for large round cells $R_0>\sqrt{\kappa_b/\gamma_c}$), 
cell division occurs for values of $\omega \sim \gamma$, corresponding to
$W_a \sim \gamma_c \sim 5\cdot 10^{-3}$~J/m$^2$.
 This value, significantly higher than
 the previous estimate,
is closer to conventional values for liquid crystals~\cite{Degennes}.
In spite of the ambiguity in the estimate of the anchoring strength, an important prediction of our model is that in all cases the anchoring of filaments to the cell cortex is necessary for a successful bipolar cell division.

There is experimental evidence that mammalian cells swell prior to cell division~\cite{Zlotek-Zlotkiewicz} and that centrosomes may also change size~\cite{Conduit}. Additionally, a correlation between cell  and centrosome sizes has been demonstrated in {\it C. Elegans}~\cite{Greenan}. 
Inspired by these findings, we have examined the influence of cell and centrosome sizes on the energetics of our model. The size of the centrosomes corresponds to the core size $\epsilon$ of topological defects and enters as a small-scale cutoff in the calculation of the elastic energy. The effect of a simultaneous $50 \%$ swelling of both cell and centrosomes prior to division 
is shown in   Fig.~\ref{fig:phase-diagram}A,B (orange squares). 
 We find that when $\gamma=0$ (frame A of Fig. \ref{fig:phase-diagram}) swelling facilitates division even at large bending rigidity, while when $\kappa=0$ swelling suppresses division, with the suppression increasing with line tension.  In other words,  
 our model suggests that cells can divide more easily when the boundary tension is small and the cell can swell easily. Conversely, this suggest that those cells where division is preceded by swelling are characterized by small values of boundary tension, while the dominant contribution to the deformation energy comes from bending elasticity of the cell cortex, indicating that    $W_a \sim 10^{-9}\div 10^{-6}$~J/m$^2$ may be the more plausible  estimate.\looseness=-1
 
Finally, we have verified that when varied separately, neither the centrosome size, $\epsilon$,  nor the cell size significantly impact  cell division.

\subsection{Multipolar division}

\begin{figure}[t]
\centering
\includegraphics[width=0.95\linewidth]{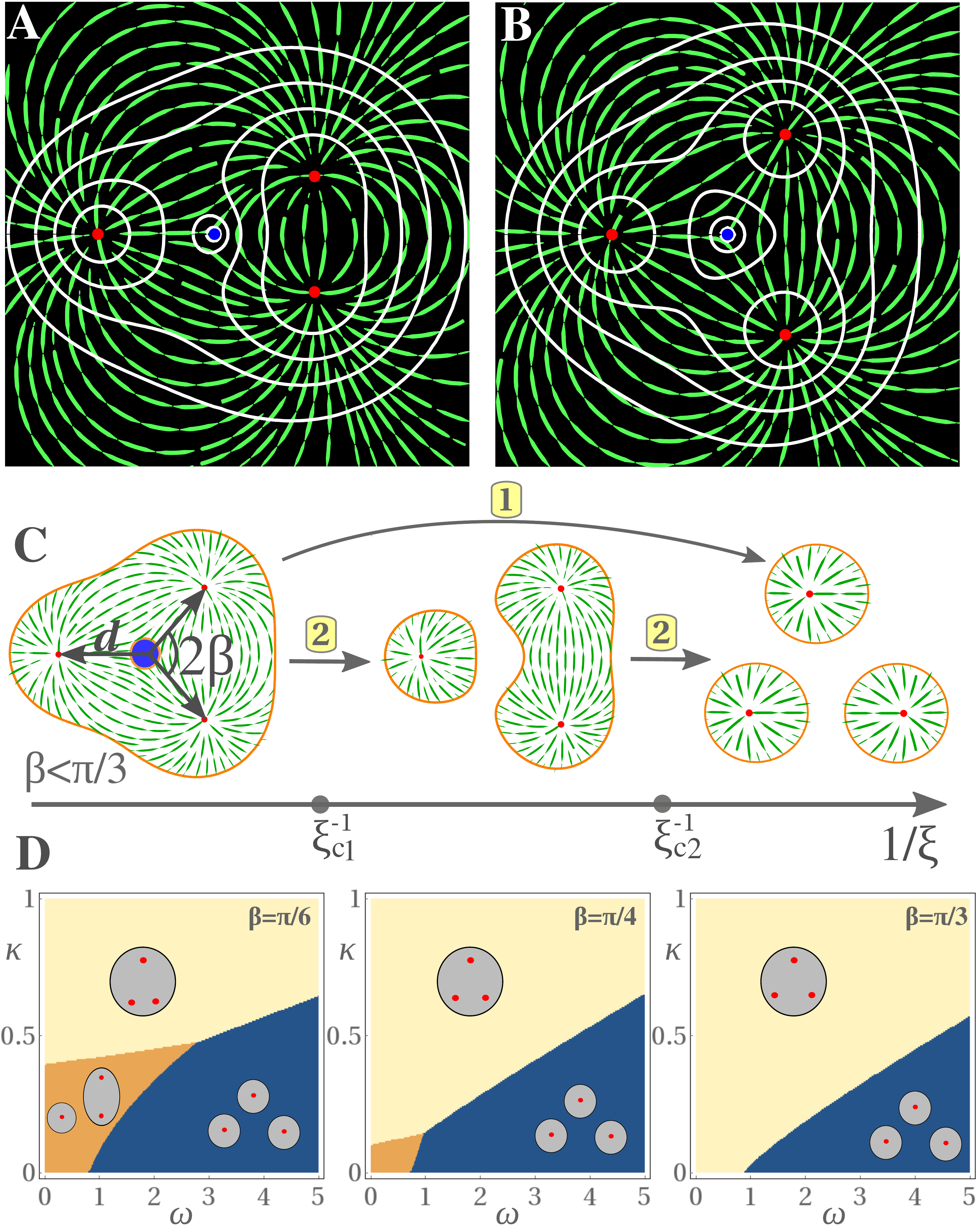}
\caption{\label{fig:field3}
Nematic defect textures and associated equi-energy boundaries for tripolar droplets with three $+1$ defects, at distance $d$ from the center,  separated by an angle~$\beta$ for $\beta=\pi/6$ (A) and $\beta =\pi/3$ (B). The color-code is the same as in Fig.~\ref{fig:field2}. Additionally,
the blue dots are  $-1/2$ disclinations of the director field, representing abnormal DNA configurations. C. Division can  occur following two distinct pathways: 1) a direct division into three  daughter droplets; or 2) a two-step division of one cell into three  daughter droplets via an intermediate structure of bipolar droplets with distinct numbers of defects.
D. Phase diagrams showing regions when the division into three (blue color) or two (orange) daughter cells is energetically favored for different initial configurations of centrosomes with opening angle $2\beta$. Bipolar division occurs for $\beta  \lesssim \pi/4$, in good agreement  with experiments on {\it Drosophila} wing disc epithelium~\cite{Sabino}, and for small values of anchoring $\omega$. The  values of the remaining parameters are $\gamma=0$ and $\epsilon = 0.01$.}
\end{figure}

\begin{figure}[t]
\centering
\includegraphics[width=0.95\linewidth]{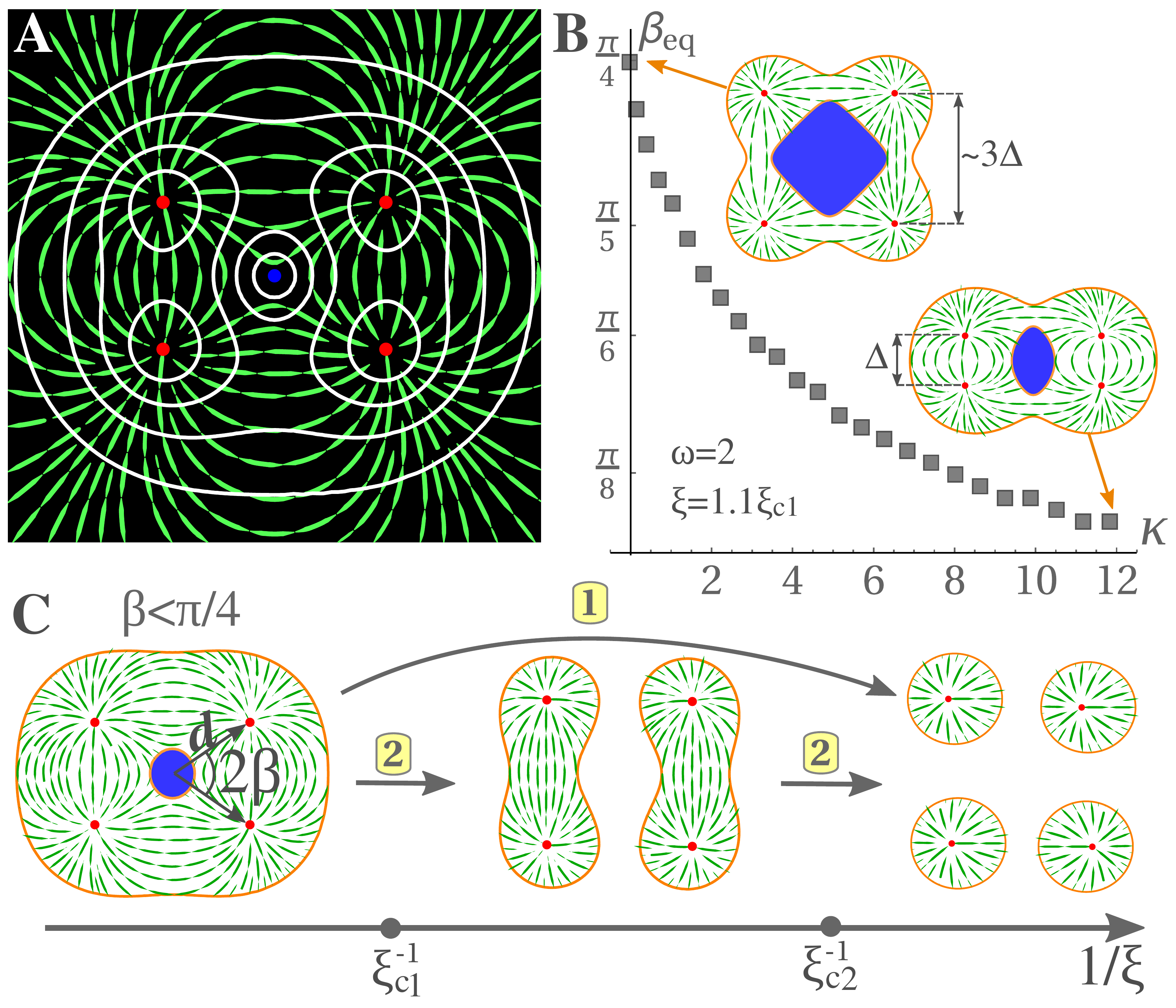}
\caption{\label{fig:field4}
A. Nematic defect textures and associated equi-energy boundaries for quadrupolar droplets with four $+1$ defects, at distance $d$ from the center and separated by angle $\beta$,  and a $-1$  disclination, representing  abnormal DNA configurations, at the center. 
The color-code is as in Fig.~\ref{fig:field2}.
B. The
 angle  $\beta_{\rm eq}$ that minimizes the  free energy~\eqref{eq:ftot} 
before division
 for the symmetric quadrupolar configuration as a function of the rescaled bending rigidity, $\kappa$.
 $\beta_{\rm eq}$ decreases with increasing $\kappa$, resembling centrosome clustering~\cite{Marthiens}. Studying this phenomenon may elucidate the physical mechanisms governing clustering in multipolar spindles.
C. Division can occur following two distinct pathways: 1) directly into four identical cells; or 2) indirectly into four daughter droplets, through an intermediate state of two bipolar droplets with equal numbers of defects.
} 
\end{figure}

Cells are sometimes observed to divide abnormally into more than two daughter cells (multipolar cell division). This can be
a consequence of genetic modifications,
controlling the expression levels of proteins that are important for mitosis,
 or  of the failure  of a previous cell division.
 What drives such faulty divisions is not understood, although accumulation of extra  
 centrosomes  has been suggested as a possible cause~\cite{Brinkley,Duncan,godinho}.
 Our model allows us to analyze configurations of microtubules corresponding to multipolar spindles by modeling them as nematic droplets with more complex defect structures, consisting of both positively charged defects representing centrosomes and negatively charged defects representing 
 abnormal DNA configurations which could be due to 
abnormal DNA content~\cite{Duncan,schatten:2008},
 shown as red and blue dots, respectively, in Figs.~\ref{fig:field3} 
 and ~\ref{fig:field4}. 
 Combining positive and negative strength defects one can design defect textures that closely resembles the configuration of microtubules inside cells with either 
tripolar~\cite{Sabino,Duncan}, Fig.~\ref{fig:spindle}B, or quadrupolar~\cite{schatten:2008,Duncan} spindles. 
The size and  orientation of symmetric tripolar and quadrupolar defect configurations can be characterized by  the distance $d$ from the center  of each of the $+1$ disclinations and by their angular separation $\beta$, as shown in Figs.~\ref{fig:field3}C, \ref{fig:field4}C.  The negative $-1/2$ disclination in the case of tripolar spindle (Fig.~\ref{fig:field3}A,B) and the $-1$ defect for the quadrupolar spindle (Fig.~\ref{fig:field4}A) account for the absence of nematic order at the center of the droplet,  or abnormal DNA configurations
at the center of these faulty dividing cells. 
This interpretation is also supported by other experimental works~\cite{Dinarina}
 showing that the DNA can act as a nucleator of microtubules and can affect multipolar spindle organization.\looseness=-1

In a tripolar spindle, two distinct pathways of cell division arise for small values of the centrosome angular separation $\beta < \pi/3$. A tripolar droplet with a hole in the nematic texture at its center can either split into two domains with unequal number of centrosomes or into three equal domains as illustrated in Fig.~\ref{fig:field3}C. When 
$\beta \geqslant \pi/3$ 
the only possible transition is from one into three cells  (see Fig.~\ref{fig:field3}C). Similarly, the division of a quadrupolar droplet can also follow two pathways, as illustrated in  Fig.~\ref{fig:field4}C.  We conjecture again that the selection of one of the two possible pathways is controlled by energetics according to the same rules we introduced for the bipolar case.  
For symmetric multipolar spindles 
 cell shapes depend at least on two parameters, $\xi$ and the angle $\beta$ defined in 
Figs.~\ref{fig:field3}C,~\ref{fig:field4}C.
There are two critical lines $\xi_{c1}(\beta)$ and $\xi_{c2}(\beta)$ that control the transitions from single connected domains to two daughter cells and from two to three or four cells, respectively, as shown in Figs.~\ref{fig:field3}C,~\ref{fig:field4}C  and summarized in Fig.~\ref{fig:3top} in Appendix.
\looseness=-1

As discussed above, the line tension $\gamma$ does not play an important role in the energetics,  thus  it will be neglected  for tripolar spindles. The regions of stability of dividing tripolar spindles are shown in  Fig.~\ref{fig:field3}D. Note that a small value of $\beta$ promotes cell division, and specifically promotes fission into  two or three daughter cells.
An important difference as
compared to the case of bipolar spindles (see Fig.~\ref{fig:phase-diagram}), is that here for small angles $\beta$, 
division can occur also in the absence of anchoring, $\omega = 0$. 
This
suggests that in multipolar spindles  cells may regulate the strength of $\omega$ depending on the angular separation  $\beta$ to ensure the division process.
Finally, for $\beta\gtrsim\pi/4$ the division of the tripolar spindle into two daughter cells is  strongly suppressed as compared to division in three cells, and does not occur for $\beta\geqslant \pi/3$. This result is in agreement with  the experimental observation of cell division in populations of {\it Drosophila} epithelial cells {\it in  vivo}, including quantitative agreement for the value of the initial angle $\beta$ suppressing dipolar division~\cite{Sabino}.

\section{Discussion and concluding remarks}
Growth and division are the hallmark of living systems, but related processes are observed also in inert matter.  
For instance,  atomic nuclei can grow by capturing neutrons~\cite{Ageno}
and then
divide into two lighter daughter nuclei via a fission process that releases energy and has been modeled 
using an
analogy with liquid droplets~\cite{Bohr}. In living cells division is driven by active processes that underlie the assembly and reorganization of the spindle, a microtubule structure that resembles a  tactoid -- an elongated nematic liquid crystal droplet with two sharp poles, with shape being mainly controlled surface tension anisotropy (strong tangential anchoring)~\cite{virus:1941,Kim}.  
Inspired by these ideas,  
we have studied
 two-dimensional nematic droplets bound by a membrane with finite bending rigidity and line tension
 to gain insight on the interplay between 
spindle elasticity and the properties of the bounding cell cortex 
 in  dividing cells.  
 While
 the importance of nematic alignment of microtubules in controlling the organization of the spindle has been demonstrated before~\cite{Brugues}, new elements of our model are the proposed connection between centrosomes and the defects that are required by topology in nematic liquid crystals and their role in controlling shape changes.  
 Thanks to this powerful analogy,
 we investigate analytically the interplay of  bulk and surface elasticity, including the role of anchoring of microtubules to the cell boundary, in controlling cell shape. 
 Our model applies provided the cell cortical boundary is deformable and there is a strong coupling between its shape and microtubule organization, as {may be} the case for  epithelial animal cells.
 It does not, however, apply  to  plant cells  that have  rigid cell-walls whereby  cells grow thanks to the help of turgor pressure~\cite{plant:2014}. 
 
 {
{ {Our} suggestion that cell division is driven entirely by the {direct} mechanical deformations that the spindle induces on the cortex appears in contrast with the accepted view that the spindle initiates signaling pathways that control the localization and activity of different molecules on the cortex, which in turn cause the shape changes associated with cell division~\cite{Pollard,Eggert,Green,Fededa}. It should be stressed, however, that our work describes a coarse-grained model where biochemical signaling processes, although not explicitly included, should be thought of as determining the effective parameters of the model, such as the shape parameter $\xi$ that controls the separation of the spindle's poles and the parameters $\kappa$, $\gamma$ and $\omega$ controlling the mechanical properties of the cortex and the strength of anchoring. Our {work}  could therefore be reconciled with the current literature for instance by assuming that the cortex properties $\kappa$, $\gamma$ and $\omega$  depend on the shape parameter $\xi$, or are more  generally 
{ time-dependent and varying  along the cortex contour} to mimic how 
spindle initiates signaling pathways.
 } 
  }
 
Our work makes several predictions that 
emphasize
the key role of centrosomes in driving both healthy bipolar and faulty multipolar division processes. The first important new result is that anchoring is always required for successful bipolar cell division. This prediction could be tested experimentally by regulating the level of expression of  proteins (e.g., of the EB family) responsible for connecting  microtubules to the cell cortex. Alternatively, one may control the anchoring by mechanically intervening on such connections, 
for instance with laser ablation~\cite{Khodjakov}. 
Our theory also provides an estimate of the anchoring strength   as $W_a \sim 10^{-9} \div 10^{-6}$~J/m$^2$, lower than typical values for liquid crystals. 
Our second prediction concerns the role of the boundary elasticity in controlling the fate of cell division, captured in our continuum model by two  parameters: the bending rigidity $\kappa$ and the tension $\gamma$. The bending rigidity  $\kappa$ of the cortical boundary could be  modulated by varying acto-myosin activity for instance via the addition of commercially available drugs such as blebbistatin. The boundary tension  $\gamma$ may perhaps be varied by acting on the physical properties of the culture medium surrounding the cells (e.g., its viscosity). Experiments of this type would provide a direct connection between the continuum mechanics of cells and various subcellular signaling processes.
Thirdly, for tripolar  spindles, we find that there is a critical angle characterizing centrosome organisation beyond which division into three daughter cells is more likely than bipolar division. 
Remarkably, a similar phenomenon was observed in {\it Drosophila} epithelial cells~\cite{Sabino}. 
The critical angle predicted by our theory is 
in good agreement with the experimental measurements.
Finally, we have demonstrated  that the combined effect of cell and centrosome size change (swelling) can help division, allowing it to occur in regions of the parameter space which are unaccessible in the absence of swelling.\looseness=-1

Many open questions lie ahead. 
An important challenge is to relate the shape parameter $\xi$ to active processes in the cell. This will require starting from a more microscopic model that incorporates dynamical processes such as microtubule polymerization and depolimerization as well as the motor's binding/unbinding and then obtain mesoscopic equations via  coarse-graining procedures. 
This is a challenging open problem that remains beyond the scope of the present work.
It would also be interesting  to extend the model to $3D$, which may be relevant for other types of cells, e.g., sea urchins~\cite{schatten:2008}.
Our model could  also be adapted to investigate  centrosome clustering~\cite{Marthiens}.
{Our preliminary analysis (see Fig.~\ref{fig:field4}B) suggests that bending rigidity  $\kappa$  and surface tension $\gamma$ 
can play an important role in controlling the angular separation among defects, hence their clustering. 
{ A} comprehensive analysis of defect clustering,
when the defect separation becomes comparable to their { core} size,
 requires { a more detailed description of the core region~\cite{Degennes} and is beyond the scope of the present work.}
}
Finally, our results also apply to synthetic active nematic droplets and reconstituted in-vitro systems
where orientational order, anchoring of filaments to the boundaries,  and boundary elasticity may be designed and tuned  more easily than in biological cells.\looseness=-1





\section{Appendix}

\subsection{Variational problem: boundary condition}
\label{sec:Appendix-boundary-condition}
We consider a nematic liquid crystals described by the director $\bn=\cos\alpha\bi_x+\sin\alpha\bi_y$ confined to a domain $\Sigma$ in 2D, and enclosed by the boundary  $\p\Sigma$ with normal  $\pmb \nu$, with free energy given by~\eqref{eq:ftot}. The equilibrium configuration of $\bn$ and the associated boundary condition can be found by setting to zero the first variation of the free energy~\eqref{eq:ftot} with respect to the angle $\alpha$, yielding~\cite{SM:2015}
\be\label{eq:bc}
\nabla^2\alpha=0, \quad \left[ 
K\,{\pmb\nu}\cdot\nabla\alpha +W_a  \sin(\alpha-\vp)\cos(\alpha-\vp)\right]_{\p\Sigma}=0,
\ee
where $\vp$ is the angle of $\pmb \nu$ with the $x$-axis. 
In  order to proceed with analytical methods, we first neglect the anchoring term, so that~\eqref{eq:bc} becomes
\be\label{eq:bc-free}
\nabla^2\alpha=0, \quad {\pmb \nu\cdot\nabla \alpha|_{\p\Sigma}=0},
\ee
which corresponds to stress free boundary condition~\cite{Landau6}. The solution of \eqref{eq:bc-free}  will yield a family of cell shapes. The physical solution will then be selected as the one that minimizes the free energy~\eqref{eq:ftot}, including the anchoring energy. 

It is convenient to work with complex coordinates where~\eqref{eq:bc-free} becomes 
\be\label{eq:bc-complex}
\p_z\p_{\bar z}\alpha=0, \quad {\left[\p_z\alpha e^{i\vp}+\p_{\bar z}\alpha e^{-i\vp}\right]_{\p\Sigma}=0},
\ee
with $z = x + i y$ and $\bar z=x-iy$.
The general solution  $\alpha$ can then be written in terms of the potential $\Omega(z)=\Pi_k(z-z_k)^{q_k}$ as (see~\eqref{eq:bulk} in the main text)
\be\label{eqs:bulk}
\alpha=\frac{\log\Omega(z)-\log\Omega(\bar z)}{2i}\;,  
\ee 
which describes configurations of the director field with defects  of strength   $q_k$ located at position $z_k$, where the nematic order vanishes. Here $q_k$ is  the winding number of the director around any closed contour surrounding the defect 
\be\label{eq:charge}
q_i=\frac 1{2\pi}\oint ds\,(\bn\times\p_s\bn)=\frac 1{2\pi}\oint d\alpha.
\ee

\subsection{Solution for cell boundaries}
\label{sec:Appendix-boundary-solution}
Given the bulk solution for $\alpha$ given by~\eqref{eqs:bulk}, we  extend it up to the boundary  $\p\Sigma$ described by a unit normal $\pmb\nu$ that satisfies the boundary condition $\pmb\nu\cdot\nabla\alpha|_{\p\Sigma}=0$ as follows.
First, we parametrize the boundary $\p\Sigma$ via a curvilinear coordinate $s$ and write
  $\pmb\nu=\cos\vp(s)\,\bi_x+\sin\vp(s)\,\bi_y$.
  Next, we assume that
the boundary can be parametrized in the complex plane by  a Schwarz function  $\bar z\equiv g(z)$,
obtained by solving $f( \frac{z+\bar z}2, \frac{z-\bar z}{2i}) = 0$ for $\bar z$. This can be related to the angle $\vp$  as (see e.g.~\cite{benamar,kera})
\be\label{eq:g}
\quad \frac{d\bar z}{\p s}=g' \frac{dz}{ds}=-i e^{-i\vp}, \quad
 \to \quad \frac{d\bar z}{dz}=g'=-e^{-2i\vp}.
\ee 
Substituting the angle $\alpha$~\eqref{eqs:bulk} into~\eqref{eq:bc-complex} and making use of~\eqref{eq:g}, the stress free boundary condition~\eqref{eq:bc} corresponding to $W_a\to0$, becomes 
\be\label{eq:gWa0}
g'=\frac{\p_z\alpha}{\p_{\bar z}\alpha}\quad \to \quad -\left[\frac{\p_z\Omega(z)}{\Omega(z)}\cdot\frac{\Omega(\bar z)}{\p_{\bar z}\Omega(\bar z)}\right]_{\p\Sigma}=\frac {d\bar z}{dz}\bigg|_{\p\Sigma}\;,
\ee
or
\be\label{eq:level}
\left[\log(\Omega(z))+\log(\Omega(\bar z))\right]_{\p\Sigma}=\const.
\ee

It follows from \eqref{eq:level} that the function $\mathcal{F}(z,\bar{z};\{z_k\})=\Omega(z)\Omega(\bar z)$ is constant on the boundary. It is then convenient to parametrize the family of allowed shapes corresponding to a given defect configuration in terms of a dimensionless
 parameter $\xi$ (level set) setting
\begin{equation}
\left[\mathcal{F}\left(\frac{z}{d},\frac{\bar{z}}{d};\big\{\frac{z_k}{d}\big\}\right)\right]_{\p\Sigma}=\xi,
\label{eqs:xi}
\end{equation}
where $d$ measures the distance of the defects from the center of the symmetric defect configuration.  
More precisely, for bipolar spindles, where $z_k=\pm d$ (Fig.~\ref{fig:field2}A), $2d$ is the separation of the two $+1$ defects. For tripolar and quadrupolar spindles, with $z_k=\{0,-d,d e^{\pm i\beta}\}$ (tripolar, Fig.~\ref{fig:field3}A,B) and $z_k=\{0,d e^{\pm i\beta},de^{i(\pi\pm\beta)}\}$ (quadrupolar, Fig.~\ref{fig:field4}A), an additional angle $\beta$ is needed to describe the defect configuration.  Note that the resulting shapes~\eqref{eqs:xi} are smooth functions for all values of $\xi$, except for the critical values $\xi_c$ where the pinching-off of the cell into daughter cells occurs. This is the result of the chosen boundary condition~\eqref{eq:bc-free}, which depends on $\nabla\alpha$. In the opposite limit,  when the line tension  anisotropy (anchoring) is the dominant term in~\eqref{eq:bc}, one would expect to find shapes with cusps, also known in the liquid crystal literature as tactoids~\cite{virus:1941}. In this limit tactoid-type shapes with cusps and the director field tangential everywhere to the boundary can be obtained either solving for the Schwarz function $g$ as above  or using the Wulff construction as in~\cite{Kim,schoot}.  
{Interfacial cusp formation,  followed by the nucleation of two +1/2 disclinations, was found within the Landau-deGennes theory during isotropic-nematic phase transition with homeotropic anchoring at the interface~\cite{rey}.
}
\looseness=-1

\subsection{Incompressibility condition}
\label{sec:Appendix-incompressibility}
Eq.~\eqref{eqs:xi} parametrizes the family of boundaries shown as white contours in~Figs.~\ref{fig:field2}A, \ref{fig:field3}A-B,~\ref{fig:field4}A, for a given value of $d$.  To compute the actual droplet  shapes (see Figs.~\ref{fig:field2}C, \ref{fig:field3}C, \ref{fig:field4}C), we assume that the droplet is incompressible, i.e., that its area is constant. 
In the case of bipolar division the areas before ($\xi>1$) and after ($\xi<1$) division, are  given respectively, by 
\begin{align}
\label{eq:s1}
S_1&=\int_{-\pi}^{\pi}\!\!\!\!d\theta \int_0^{r_1}\!\!\!\!dr\,r=d^2 \bigg[\sqrt{\xi -1} {\cal E}\bigg(\frac{1}{1-\xi }\bigg)+\sqrt{\xi }{\cal E}\bigg(\frac{1}{\xi }\bigg)\bigg],\\
\label{eq:s2}
S_2&=2\int_{-\theta_m}^{\theta_m}\!\!\!d\theta \int_{r_2}^{r_1}\!\!\!dr\,r=2d^2 \sqrt{\xi }{\cal E}\bigg(\arcsin\sqrt{\xi }\bigg|\frac{1}{\xi }\bigg),
\end{align}
where ${\cal E}$ is the complete elliptic integral of the second kind. Assuming  that $S_1=\pi R_0^2=S_2$, the distance $d$ between two defects~in\eqref{eq:level2} becomes a function of $\xi$. Thus in the two regions, before and after division, $d$ can be expressed as two distinct functions of $\xi$ which match at $\xi_c=1$.

\subsection{Topological transformations}
Here we discuss  how $\xi$  controls the change of topology that accompanies cell division. 

\begin{figure}[H]
\centering
\includegraphics[width=.9\linewidth]{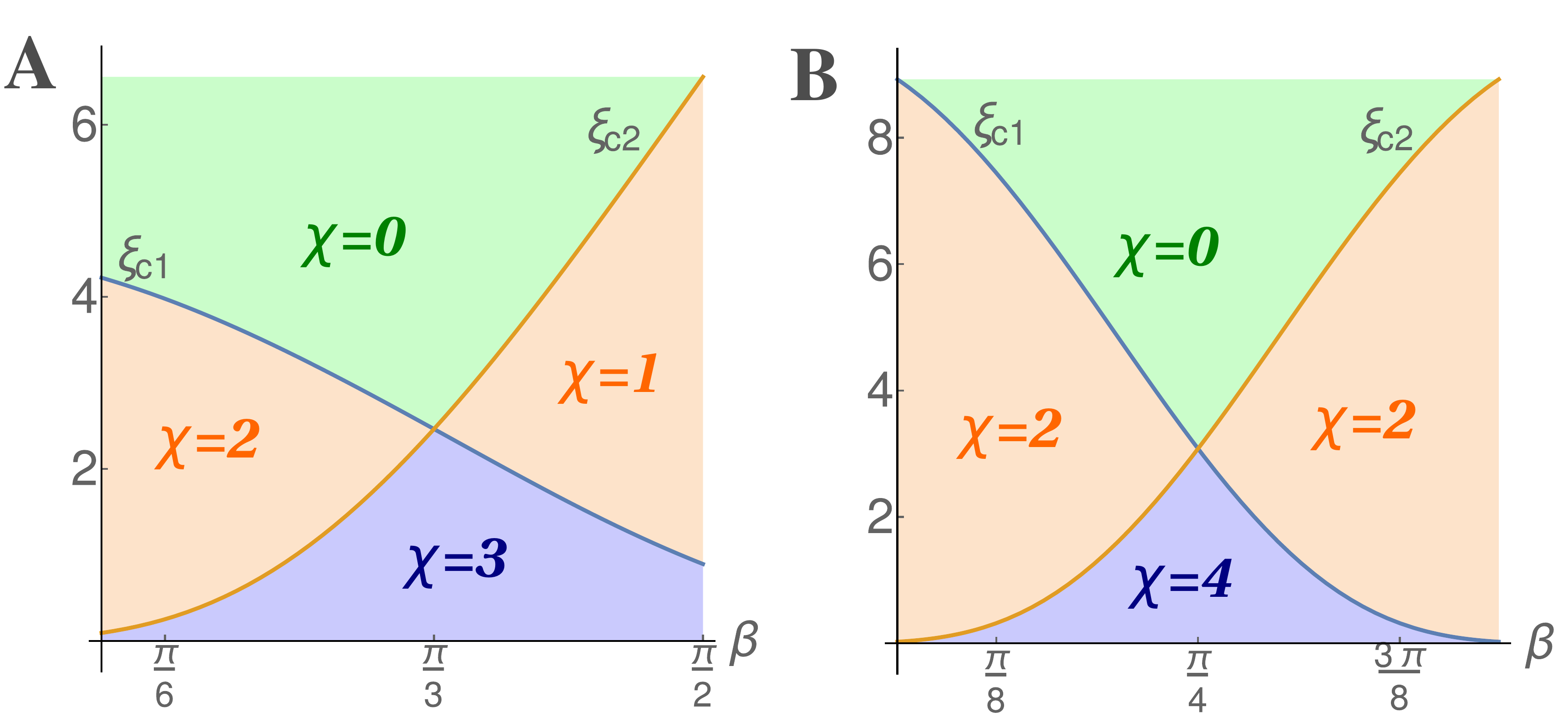}
\caption{\label{fig:3top}The regions relating the discrete Euler characteristic $\chi$, with the continuous variation of the parameter $\xi$ in case of tripolar (A) and quadrupolar (B) division. The angle $\beta$ describes the chosen configuration of $+1$ defects, with symmetry of the curves~\eqref{eq:critxi} around $\beta=\pi/4$ (B).}
\end{figure}

For a tripolar spindle, described as three $+1$ disclinations surrounding  a single $-1/2$ disclination, the equation for the boundary~\eqref{eqs:xi} can be written in polar coordinates as \begin{multline}
\label{eq:level3}
(r^2+d^2+2r d\cos\theta)(r^2+d^2-2r d\cos(\theta-\beta))\times\\\times(r^2+d^2-2r d\cos(\theta+\beta))=\xi r d^5.
\end{multline}
Unlike~\eqref{eq:level2} for the bipolar case, there is no closed form solution for the radius vector. Using {\it Mathematica} we can express implicitly $r(d,\xi,\beta,\theta)$ and analyze a variety of contours for different angles $\beta$ and level sets~$\xi$. More important, we identify the extrema of the function~\eqref{eq:level3} and the   
critical values $\xi_{c1}$ and $\xi_{c2}$ where the topological transitions occur~(Fig.~\ref{fig:3top}A) at $\theta=0$, 
$\theta|_{\beta<\pi/3}=\arccos[-\sqrt{(2\cos\beta-\cos(2\beta))/6}]$, $\theta|_{\beta>\pi/3}=\arccos[(\cos\beta-2)/3]$. 

For a quadrupolar spindle, described as an arrangement of four $+1$ disclinations surrounding a $-1$  disclination, the equation for the boundary~\eqref{eqs:xi} can be written in polar coordinates as
\begin{multline}\label{eq:level4}
2 r^4 d^4 \cos (4 \beta )+2 r^4 d^4 \cos (4 \theta)+\\+(r^4+d^4) (r^4-4 r^2 d^2 \cos (2 \beta ) \cos (2 \theta)+d^4)=\xi r^2 d^6.
\end{multline}
The critical values of $\xi$ when a single domain splits into two and four domains (Fig.~\ref{fig:3top}B) at $\theta=\{0,\pi/2\}$ are
\be\label{eq:critxi}
\xi_{c1,c2}=\frac{4}{27} \big[\pm 33 \cos (2 \beta )\mp \cos (6 \beta )+\sqrt{2} (7+\cos (4 \beta ))^{3/2}\big].
\ee

There is a  mathematical relation between the sum of the charges~\eqref{eq:charge} of the defects in the nematic texture, the Euler characteristic $\chi$~\cite{doCarmo} of the confining boundary and the integral of the angle deficit between the director $\bn$ and the 
normal $\pmb\nu$ along the boundary  $\p\Sigma$, given by~\cite{SM:2015}
\be\label{eq:chi}
\sum_i q_i
-
\frac 1{2\pi}\int_{\p\Sigma}d(\alpha-\vp)
=\chi, \quad \bn\cdot\pmb\nu=\cos(\alpha-\vp).
\ee
In 2D $\chi$ is related to the number ${\cal H}$ of holes in the nematic texture via $\chi=1-{\cal H}$. Then for a  disc $\chi=1$ and $\chi=0$ for an annulus. The Euler characteristic $\chi$ can therefore be interpreted as  identifying the number of daughter cells after division. Moreover, the director field is on average aligned along the normal to the boundary  $\p\Sigma$, for daughter cells with only one $+1$ disclination (Fig.~\ref{fig:3top}A-D), as imposed by the conservation law~\eqref{eq:chi}.  In other cases, considered above where $\sum_i q_i>1$, the director $\bn$ and the normal $\pmb \nu$ are not aligned, resulting into a non-zero contribution to the integral at the left hand side of the~\eqref{eq:chi}. We have shown above that such misalignment is important for triggering cell division.

\subsection{Method for obtaining the 
solution of  the boundary condition including anchoring,  Eq.~\eqref{eq:bc} }
 \label{sec:Appendix-perturbative}
In the preceeding section and in the main text we  computed
the exact solution, for the shape of the boundary using the simplified  
boundary  condition~\eqref{eq:bc-free}, which corresponds to
the limit of vanishingly small anchoring, 
$W_a \to 0$.
On the other hand, we have shown that the selection of 
the 
equilibrium shape itself depends on the value of $W_a$.  
In this subsection we present a method to
study systematically the effect of the boundary condition corresponding to a finite value of
the anchoring parameter $W_a$ on the shape of 
 the boundary, in agreement with previous work~\cite{protein}. 
This provides a test of
the consistency of the approximate scheme used in the main text.
To this end we first
 rewrite the boundary condition~\eqref{eq:bc} in the complex plane as
\be\label{eq:bc1}
e^{i\vp}\p_z\alpha+e^{-i\vp}\p_{\bar z}\alpha+\varpi \frac{e^{-2i\vp}e^{2i\alpha}-e^{2i\vp} e^{-2i\alpha}}{2i}=0\;,
\ee
where the complex coordinates $z,\bar z$ are scaled by the distance $d$~\footnote{Alternatively one could
use the typical size of the domain $R_0$. 
Here these two sizes are related through the incompressibility condition of the domain, $d(R_0)$.} and  we have introduced the dimensionless parameter $\varpi\equiv{W_a d}/{2K}$. 
We can then solve~\eqref{eq:bc1} for small  $\varpi$  by expanding the Schwarz function $g$~\eqref{eq:g} as a power series in $\varpi$, 
\be\label{eq:gseries}
 g=g_0+\varpi g_1+\varpi^2g_2\ldots.
\ee

\begin{figure}[h]
\centering
$\xi=4$\hskip22mm $\xi=1.2$\hskip22mm$\xi=0.9$\\[1ex]
\includegraphics[width=25mm]{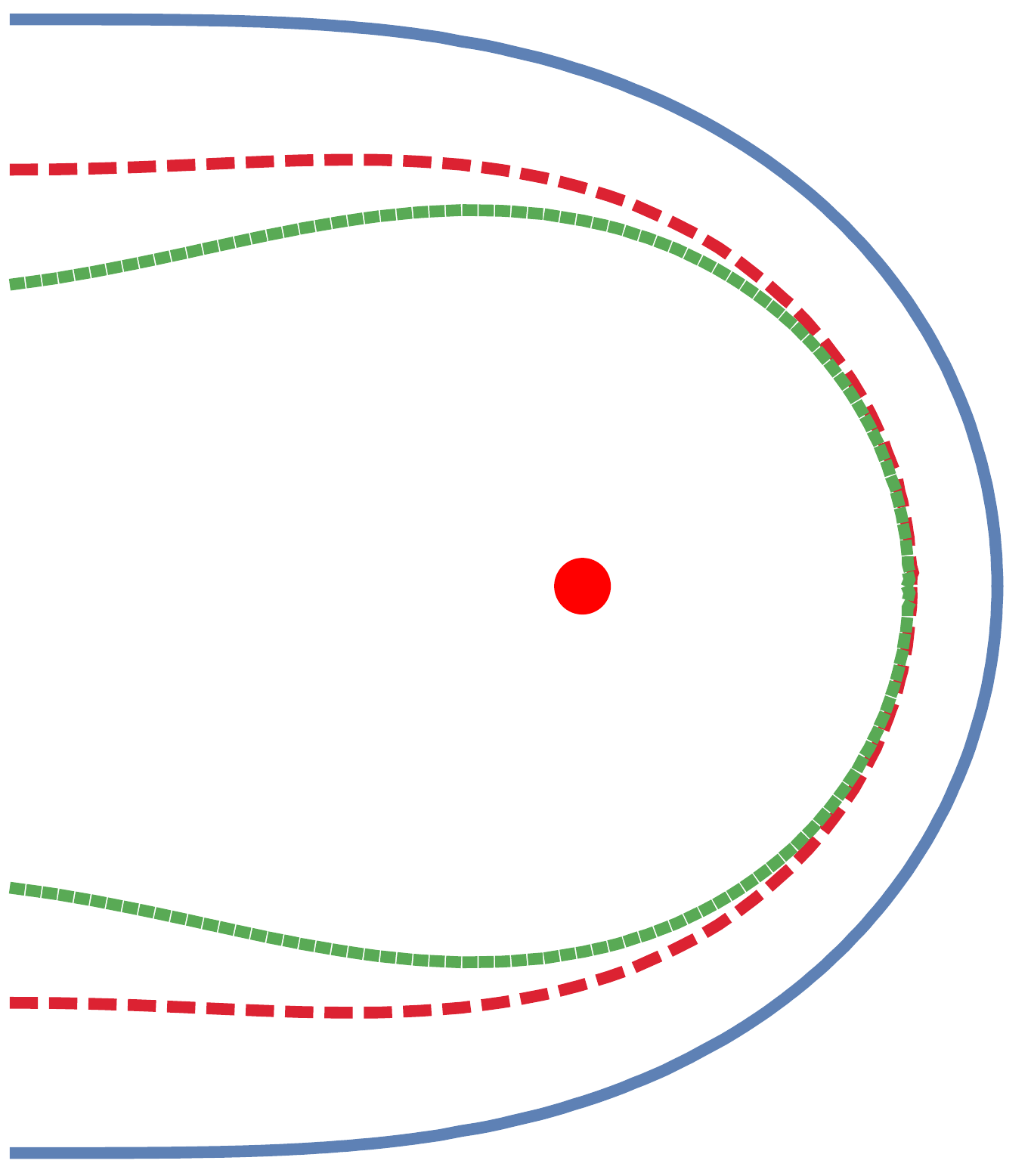}\hfill
\includegraphics[width=30mm]{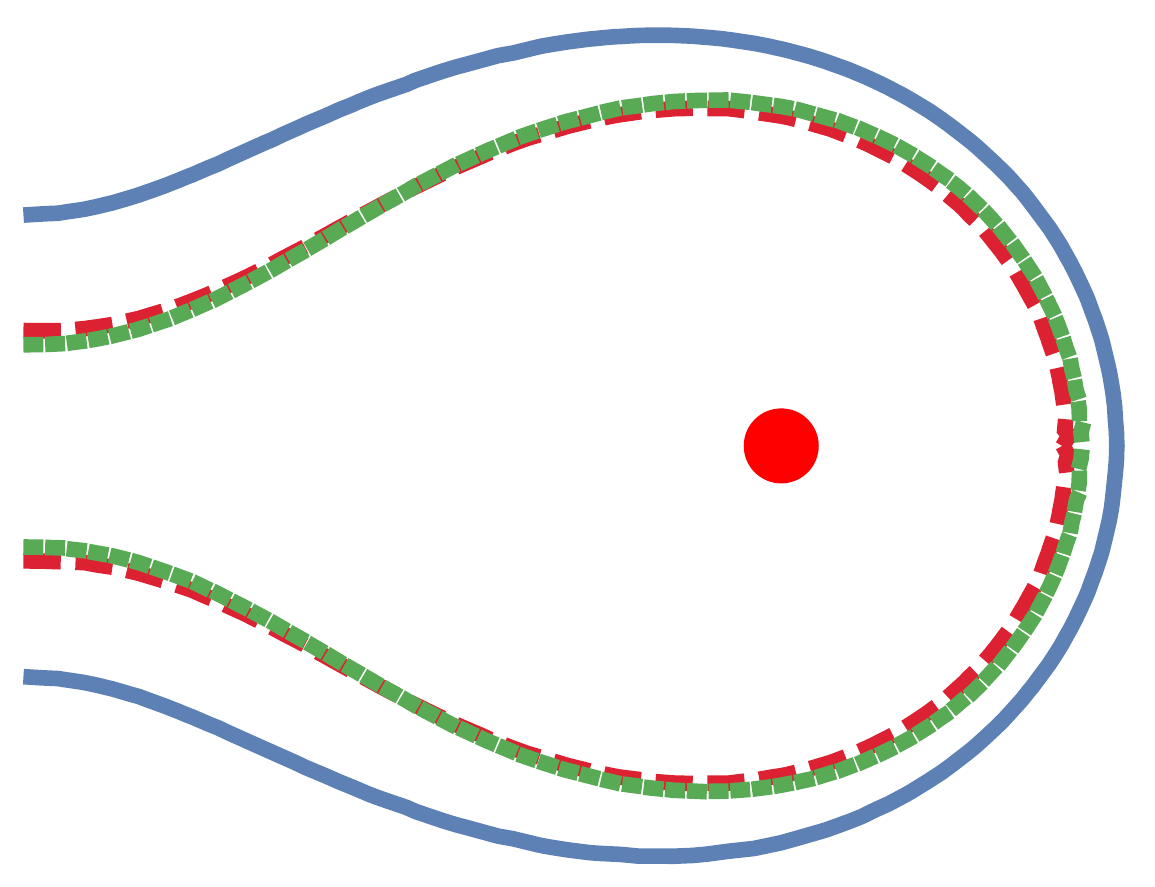}\hfill
\includegraphics[width=30mm]{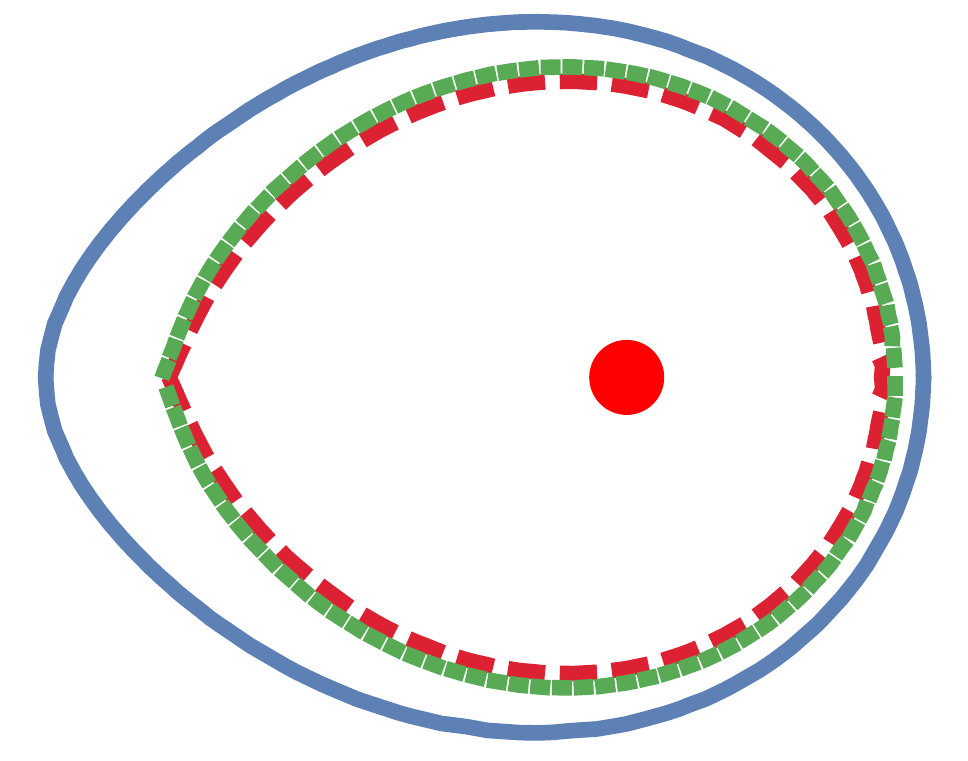}
\caption{
Comparison of the boundary contours 
obtained via various approximations: without anchoring, 
as in the main text, 
(blue curve, $\bar z=g_0$) and with $\varpi=0.4$ (red and green curves,  $\bar z=g_0+\varpi g_1$). We show the solution  of~\eqref{eq:bc1} only around the $+1$ disclination, 
since the solutions have
mirror symmetry with respect to the vertical axis, so that blue curves here coincide with the right
portions of the contours shown in Fig.~\ref{fig:field2}. 
The three panels are for 
$\xi = 4$, $\xi = 1.2$ and $\xi = 0.9$, 
as indicated.
The correction $g_1$ due to anchoring is 
evaluated including terms up to $i=2$ (red curve) and  $i=6$ (green curve) in Eq.~\eqref{eq:g1s}. 
}
\label{fig:contours}
\end{figure}

For simplicity, let us consider the case of bipolar division, with $\alpha=\Im \log (z^2-d^2)$. Substituting~\eqref{eq:gseries}  into~\eqref{eq:bc1} and including  terms 
up to first order in $\varpi$, we solve to each order, with the result
\begin{align}
O(\varpi^0):  \quad g_0&=d\sqrt{1+\frac{\xi }{(z/d)^2-1}}, \label{eq:g0}\\
O(\varpi^1): \quad g_1'&=\bigg(\frac{e^{2i\alpha} g_0'}{2i}-\frac{e^{-2i\alpha} }{2i g_0'} \bigg) \frac{g_0'}{\p_{\bar z}\alpha\sqrt{-g_0'}-\p_z\alpha/\sqrt{-g_0'}}.\label{eq:g1}
\end{align}
The zero-th order solution,
$g_0$,  is the solution of Eq.~\eqref{eq:gWa0} 
and is the same solution
plotted in Fig.~\ref{fig:field2} and expressed in  polar coordinates in Eq.~\eqref{eq:level2} of the main text. 
The correction $g_1$ 
accounts for the leading order effect of anchoring on modifying the shape of the solution. To obtain an explicit expression for $g_1$ we write it as
a power series,  
 around the position of the defect core at $z=d$, as
\be\label{eq:g1s}
g_1=\sum_{i=0}^{\infty} a_i \bigg(\frac z d-1\bigg)^{\frac{2i-1}4}.
\ee
The coefficients $a_i$ of the power series are fixed
by matching term by term the left-hand-side
and the right-hand-side of the Eq.~\eqref{eq:g1}, using  the known form of $g_0$. 
This yields  
\begin{gather}
a_0=-\frac{\xi^{3/4}}{2^{1/4}},\quad a_1=0,\quad a_2=\frac{3\xi-4}{8 (2\xi)^{1/4}}, \quad a_3=-\frac{\xi^{1/4}}{5\cdot 2 ^{3/4}},\\ a_4=\frac{336-\xi(312+11\xi)}{896\cdot(2\xi^5)^{1/4}},\quad a_5=\frac{4-7\xi}{72 (2\xi)^{3/4}},\quad\ldots
\end{gather}
In  Fig.~\ref{fig:contours} we compare the contour $\bar z=g_0$ and 
the contours $\bar z=g_0+\varpi g_1$, obtained following this procedure,
including the first two (red) and six (green)  terms in the power series~\eqref{eq:g1s}. 
We have verified that including higher order terms in the expansion of $g_1$  does not further change the shape of the green curve, suggesting that  the power series~\eqref{eq:g1s} converges. 
Moreover, the resulting shapes with finite anchoring, $\varpi=0.4$,  closely resemble
the shapes computed in the main text with the simplified boundary condition (no anchoring). 
We conclude
that  the 
 approach proposed here  and summarized in Fig.~\ref{fig:contours}
is consistent with the 
  results presented in the main text
showing that anchoring is not relevant in determining the shapes
but it plays a key role in controlling the selection of the shape 
according to our energetic criterion.

\subsection*{Acknowledgements}
{M. L. acknowledges financial
support by the ICAM Branch Contributions and stimulating discussions with Maddalena Nano. MCM was supported by the Simons Foundation through a Targeted Grant in the Mathematical Modeling of Living Systems, award No. 342354, and by the National Science Foundation through awards DMR-1305184 
and DMR-1609208. 
 All authors acknowledge support from the Syracuse University Soft Matter Program. 
}

\bibliography{notes_cell} %

\providecommand*{\mcitethebibliography}{\thebibliography}
\csname @ifundefined\endcsname{endmcitethebibliography}
{\let\endmcitethebibliography\endthebibliography}{}
\begin{mcitethebibliography}{45}
\providecommand*{\natexlab}[1]{#1}
\providecommand*{\mciteSetBstSublistMode}[1]{}
\providecommand*{\mciteSetBstMaxWidthForm}[2]{}
\providecommand*{\mciteBstWouldAddEndPuncttrue}
  {\def\EndOfBibitem{\unskip.}}
\providecommand*{\mciteBstWouldAddEndPunctfalse}
  {\let\EndOfBibitem\relax}
\providecommand*{\mciteSetBstMidEndSepPunct}[3]{}
\providecommand*{\mciteSetBstSublistLabelBeginEnd}[3]{}
\providecommand*{\EndOfBibitem}{}
\mciteSetBstSublistMode{f}
\mciteSetBstMaxWidthForm{subitem}
{(\emph{\alph{mcitesubitemcount}})}
\mciteSetBstSublistLabelBeginEnd{\mcitemaxwidthsubitemform\space}
{\relax}{\relax}

\bibitem[Albert \emph{et~al.}(2002)Albert\emph{et~al.}]{Albert}
B.~Albert \emph{et~al.}, \emph{The Molecular Biology of the Cell, 4th ed.},
  Garland Science, 2002\relax
\mciteBstWouldAddEndPuncttrue
\mciteSetBstMidEndSepPunct{\mcitedefaultmidpunct}
{\mcitedefaultendpunct}{\mcitedefaultseppunct}\relax
\EndOfBibitem
\bibitem[Brinkley and Goepfert(1998)]{Brinkley}
B.~R. Brinkley and T.~M. Goepfert, \emph{Cell Mot. and the Cytoskeleton}, 1998,
  \textbf{41}, 281--288\relax
\mciteBstWouldAddEndPuncttrue
\mciteSetBstMidEndSepPunct{\mcitedefaultmidpunct}
{\mcitedefaultendpunct}{\mcitedefaultseppunct}\relax
\EndOfBibitem
\bibitem[Godinho and Pellman(2014)]{godinho}
S.~A. Godinho and D.~Pellman, \emph{Phil. Trans. R. Soc. B}, 2014,
  \textbf{369}, 20130467\relax
\mciteBstWouldAddEndPuncttrue
\mciteSetBstMidEndSepPunct{\mcitedefaultmidpunct}
{\mcitedefaultendpunct}{\mcitedefaultseppunct}\relax
\EndOfBibitem
\bibitem[Stearns(2015)]{stearns:2015}
T.~Stearns, \emph{Science}, 2015, \textbf{1091}, 348\relax
\mciteBstWouldAddEndPuncttrue
\mciteSetBstMidEndSepPunct{\mcitedefaultmidpunct}
{\mcitedefaultendpunct}{\mcitedefaultseppunct}\relax
\EndOfBibitem
\bibitem[Marthiens \emph{et~al.}(2012)Marthiens, Piel, and Basto]{Marthiens}
V.~Marthiens, M.~Piel and R.~Basto, \emph{J Cell Sci}, 2012, \textbf{125},
  3281--3292\relax
\mciteBstWouldAddEndPuncttrue
\mciteSetBstMidEndSepPunct{\mcitedefaultmidpunct}
{\mcitedefaultendpunct}{\mcitedefaultseppunct}\relax
\EndOfBibitem
\bibitem[Basto \emph{et~al.}(2008)Basto, Brunk, Vinadogrova, Peel, Franz,
  Khodjakov, and Raff]{Basto-cell}
R.~Basto, K.~Brunk, T.~Vinadogrova, N.~Peel, A.~Franz, A.~Khodjakov and
  J.~Raff, \emph{Cell}, 2008, \textbf{133}, 1032--1042\relax
\mciteBstWouldAddEndPuncttrue
\mciteSetBstMidEndSepPunct{\mcitedefaultmidpunct}
{\mcitedefaultendpunct}{\mcitedefaultseppunct}\relax
\EndOfBibitem
\bibitem[Duncan \emph{et~al.}(2010)Duncan, Taylor, Hickey, Newell, Lenzi,
  Olson, Finegold, and Grompe]{Duncan}
A.~W. Duncan, M.~H. Taylor, R.~D. Hickey, A.~E.~H. Newell, M.~L. Lenzi, S.~B.
  Olson, M.~J. Finegold and M.~Grompe, \emph{Nature}, 2010, \textbf{467},
  707\relax
\mciteBstWouldAddEndPuncttrue
\mciteSetBstMidEndSepPunct{\mcitedefaultmidpunct}
{\mcitedefaultendpunct}{\mcitedefaultseppunct}\relax
\EndOfBibitem
\bibitem[Marthiens \emph{et~al.}(2013)Marthiens, Rujano, Pennetier, Tessier,
  {Paul-Gilloteaux}, and Basto]{Marthiens13}
V.~Marthiens, M.~A. Rujano, C.~Pennetier, S.~Tessier, P.~{Paul-Gilloteaux} and
  R.~Basto, \emph{Nat. Cell Biol.}, 2013, \textbf{15}, 731\relax
\mciteBstWouldAddEndPuncttrue
\mciteSetBstMidEndSepPunct{\mcitedefaultmidpunct}
{\mcitedefaultendpunct}{\mcitedefaultseppunct}\relax
\EndOfBibitem
\bibitem[Th\'ery \emph{et~al.}(2007)Th\'ery, Jim\'enez-Dalmaroni, Racine,
  Bornens, and J\"ulicher]{Thery}
M.~Th\'ery, A.~Jim\'enez-Dalmaroni, V.~Racine, M.~Bornens and F.~J\"ulicher,
  \emph{Nature}, 2007, \textbf{447}, 493--496\relax
\mciteBstWouldAddEndPuncttrue
\mciteSetBstMidEndSepPunct{\mcitedefaultmidpunct}
{\mcitedefaultendpunct}{\mcitedefaultseppunct}\relax
\EndOfBibitem
\bibitem[Brugu{\'e}s and Needleman(2014)]{Brugues}
J.~Brugu{\'e}s and D.~Needleman, \emph{Proc. Natl. Acad. Sci. USA}, 2014,
  \textbf{111}, 18496--18500\relax
\mciteBstWouldAddEndPuncttrue
\mciteSetBstMidEndSepPunct{\mcitedefaultmidpunct}
{\mcitedefaultendpunct}{\mcitedefaultseppunct}\relax
\EndOfBibitem
\bibitem[Sanchez \emph{et~al.}(2012)Sanchez, Chen, DeCamp, Heymann, and
  Dogic]{Sanchez2012}
T.~Sanchez, D.~T.~N. Chen, S.~J. DeCamp, M.~Heymann and Z.~Dogic,
  \emph{Nature}, 2012, \textbf{491}, 431--435\relax
\mciteBstWouldAddEndPuncttrue
\mciteSetBstMidEndSepPunct{\mcitedefaultmidpunct}
{\mcitedefaultendpunct}{\mcitedefaultseppunct}\relax
\EndOfBibitem
\bibitem[Giomi \emph{et~al.}(2013)Giomi, Bowick, Ma, and Marchetti]{Giomi2013}
L.~Giomi, M.~J. Bowick, X.~Ma and M.~C. Marchetti, \emph{Phys. Rev. Lett.},
  2013, \textbf{110}, 228101\relax
\mciteBstWouldAddEndPuncttrue
\mciteSetBstMidEndSepPunct{\mcitedefaultmidpunct}
{\mcitedefaultendpunct}{\mcitedefaultseppunct}\relax
\EndOfBibitem
\bibitem[Giomi and DeSimone(2014)]{Giomi}
L.~Giomi and A.~DeSimone, \emph{Phys. Rev. Lett.}, 2014, \textbf{112},
  147802\relax
\mciteBstWouldAddEndPuncttrue
\mciteSetBstMidEndSepPunct{\mcitedefaultmidpunct}
{\mcitedefaultendpunct}{\mcitedefaultseppunct}\relax
\EndOfBibitem
\bibitem[Salbreux \emph{et~al.}(2009)Salbreux, Prost, and Joanny]{Salbreux}
G.~Salbreux, J.~Prost and J.~F. Joanny, \emph{Phys. Rev. Lett.}, 2009,
  \textbf{103}, 058102\relax
\mciteBstWouldAddEndPuncttrue
\mciteSetBstMidEndSepPunct{\mcitedefaultmidpunct}
{\mcitedefaultendpunct}{\mcitedefaultseppunct}\relax
\EndOfBibitem
\bibitem[Turlier \emph{et~al.}(2014)Turlier, Audoly, Prost, and
  Joanny]{Turlier}
H.~Turlier, B.~Audoly, J.~Prost and J.~F. Joanny, \emph{Biophys. J.}, 2014,
  \textbf{106}, 114--123\relax
\mciteBstWouldAddEndPuncttrue
\mciteSetBstMidEndSepPunct{\mcitedefaultmidpunct}
{\mcitedefaultendpunct}{\mcitedefaultseppunct}\relax
\EndOfBibitem
\bibitem[Sedzinski \emph{et~al.}(2011)Sedzinski, Biro, Oswald, Tinevez,
  Salbreux, and Paluch]{Paluch}
J.~Sedzinski, M.~Biro, A.~Oswald, J.~Tinevez, G.~Salbreux and E.~Paluch,
  \emph{Nature}, 2011, \textbf{476}, 462--466\relax
\mciteBstWouldAddEndPuncttrue
\mciteSetBstMidEndSepPunct{\mcitedefaultmidpunct}
{\mcitedefaultendpunct}{\mcitedefaultseppunct}\relax
\EndOfBibitem
\bibitem[\AA~str\"om \emph{et~al.}(2010)\AA~str\"om, von Alfthan, Kumarb, and
  Karttunen]{mikko}
J.~A. \AA~str\"om, S.~von Alfthan, P.~B.~S. Kumarb and M.~Karttunen, \emph{Soft
  Matter}, 2010, \textbf{6}, 5375\relax
\mciteBstWouldAddEndPuncttrue
\mciteSetBstMidEndSepPunct{\mcitedefaultmidpunct}
{\mcitedefaultendpunct}{\mcitedefaultseppunct}\relax
\EndOfBibitem
\bibitem[Tirnauera and Bierer(2000)]{EBprotein}
J.~S. Tirnauera and B.~E. Bierer, \emph{J. Cell Biol.}, 2000, \textbf{149},
  761--766\relax
\mciteBstWouldAddEndPuncttrue
\mciteSetBstMidEndSepPunct{\mcitedefaultmidpunct}
{\mcitedefaultendpunct}{\mcitedefaultseppunct}\relax
\EndOfBibitem
\bibitem[Sabino \emph{et~al.}(2015)Sabino, Gogendeau, Gambarotto, Nano,
  Pennetier, Dingli, Arras, Loew, and Basto]{Sabino}
D.~Sabino, D.~Gogendeau, D.~Gambarotto, M.~Nano, C.~Pennetier, F.~Dingli,
  G.~Arras, D.~Loew and R.~Basto, \emph{Curr. Biol.}, 2015, \textbf{25},
  879--889\relax
\mciteBstWouldAddEndPuncttrue
\mciteSetBstMidEndSepPunct{\mcitedefaultmidpunct}
{\mcitedefaultendpunct}{\mcitedefaultseppunct}\relax
\EndOfBibitem
\bibitem[Manyuhina \emph{et~al.}(2015)Manyuhina, Lawlor, Marchetti, and
  Bowick]{SM:2015}
O.~V. Manyuhina, K.~B. Lawlor, M.~C. Marchetti and M.~J. Bowick, \emph{Soft
  Matter}, 2015, \textbf{11}, 6099\relax
\mciteBstWouldAddEndPuncttrue
\mciteSetBstMidEndSepPunct{\mcitedefaultmidpunct}
{\mcitedefaultendpunct}{\mcitedefaultseppunct}\relax
\EndOfBibitem
\bibitem[Desai and Mitchison(1997)]{MTdyn}
A.~Desai and T.~J. Mitchison, \emph{Annu. Rev. Cell Dev. Biol.}, 1997,
  \textbf{13}, 83--117\relax
\mciteBstWouldAddEndPuncttrue
\mciteSetBstMidEndSepPunct{\mcitedefaultmidpunct}
{\mcitedefaultendpunct}{\mcitedefaultseppunct}\relax
\EndOfBibitem
\bibitem[DeGennes and Prost(1993)]{Degennes}
P.~G. DeGennes and J.~Prost, \emph{The Physics of Liquid Crystals, 2nd ed.},
  Oxford University Press, 1993\relax
\mciteBstWouldAddEndPuncttrue
\mciteSetBstMidEndSepPunct{\mcitedefaultmidpunct}
{\mcitedefaultendpunct}{\mcitedefaultseppunct}\relax
\EndOfBibitem
\bibitem[Alert \emph{et~al.}(2015)Alert, Casademunt, Brugu\'es, and
  Sens]{sens:2015}
R.~Alert, J.~Casademunt, J.~Brugu\'es and P.~Sens, \emph{Biophys J.}, 2015,
  \textbf{108}, 1878--86\relax
\mciteBstWouldAddEndPuncttrue
\mciteSetBstMidEndSepPunct{\mcitedefaultmidpunct}
{\mcitedefaultendpunct}{\mcitedefaultseppunct}\relax
\EndOfBibitem
\bibitem[Zlotek-Zlotkiewicz \emph{et~al.}(2015)Zlotek-Zlotkiewicz, Monnier,
  Cappello, {Le Berre}, and Piel]{Zlotek-Zlotkiewicz}
E.~Zlotek-Zlotkiewicz, S.~Monnier, G.~Cappello, M.~{Le Berre} and M.~Piel,
  \emph{J. Cell Biol.}, 2015, \textbf{211}, 765--774\relax
\mciteBstWouldAddEndPuncttrue
\mciteSetBstMidEndSepPunct{\mcitedefaultmidpunct}
{\mcitedefaultendpunct}{\mcitedefaultseppunct}\relax
\EndOfBibitem
\bibitem[Conduit \emph{et~al.}(2010)Conduit, Brunk, Dobbelaere, Dix, Lucas, and
  Raff]{Conduit}
P.~T. Conduit, K.~Brunk, J.~Dobbelaere, C.~Dix, E.~Lucas and J.~Raff,
  \emph{Curr. Biol.}, 2010, \textbf{20}, 2178--2186\relax
\mciteBstWouldAddEndPuncttrue
\mciteSetBstMidEndSepPunct{\mcitedefaultmidpunct}
{\mcitedefaultendpunct}{\mcitedefaultseppunct}\relax
\EndOfBibitem
\bibitem[Greenan \emph{et~al.}(2010)Greenan, Brangwynne, Jaensch, Gharakhani,
  F., and Hyman]{Greenan}
G.~Greenan, C.~P. Brangwynne, S.~Jaensch, J.~Gharakhani, J.~F. and A.~Hyman,
  \emph{Curr. Biol.}, 2010, \textbf{20}, 353--358\relax
\mciteBstWouldAddEndPuncttrue
\mciteSetBstMidEndSepPunct{\mcitedefaultmidpunct}
{\mcitedefaultendpunct}{\mcitedefaultseppunct}\relax
\EndOfBibitem
\bibitem[Schatten(2008)]{schatten:2008}
H.~Schatten, \emph{Histochem. Cell Biol.}, 2008, \textbf{129}, 667\relax
\mciteBstWouldAddEndPuncttrue
\mciteSetBstMidEndSepPunct{\mcitedefaultmidpunct}
{\mcitedefaultendpunct}{\mcitedefaultseppunct}\relax
\EndOfBibitem
\bibitem[Dinarina \emph{et~al.}(2009)Dinarina, Pugieux, Corral, Loose, Spatz,
  Karsenti, and N{\'e}d{\'e}lec]{Dinarina}
A.~Dinarina, C.~Pugieux, M.~M. Corral, M.~Loose, J.~Spatz, E.~Karsenti and
  F.~N{\'e}d{\'e}lec, \emph{Cell}, 2009, \textbf{138}, 502--513\relax
\mciteBstWouldAddEndPuncttrue
\mciteSetBstMidEndSepPunct{\mcitedefaultmidpunct}
{\mcitedefaultendpunct}{\mcitedefaultseppunct}\relax
\EndOfBibitem
\bibitem[Ageno(1965)]{Ageno}
M.~Ageno, \emph{Nature}, 1965, \textbf{205}, 1307\relax
\mciteBstWouldAddEndPuncttrue
\mciteSetBstMidEndSepPunct{\mcitedefaultmidpunct}
{\mcitedefaultendpunct}{\mcitedefaultseppunct}\relax
\EndOfBibitem
\bibitem[Bohr and Wheeler(1939)]{Bohr}
N.~Bohr and J.~A. Wheeler, \emph{Phys. Rev.}, 1939, \textbf{56}, 426--450\relax
\mciteBstWouldAddEndPuncttrue
\mciteSetBstMidEndSepPunct{\mcitedefaultmidpunct}
{\mcitedefaultendpunct}{\mcitedefaultseppunct}\relax
\EndOfBibitem
\bibitem[Bernal and Fankuchen(1941)]{virus:1941}
J.~D. Bernal and I.~Fankuchen, \emph{J Gen Physiol.}, 1941, \textbf{25},
  111--146\relax
\mciteBstWouldAddEndPuncttrue
\mciteSetBstMidEndSepPunct{\mcitedefaultmidpunct}
{\mcitedefaultendpunct}{\mcitedefaultseppunct}\relax
\EndOfBibitem
\bibitem[Kim \emph{et~al.}(2013)Kim, Shiyanovskii, and Lavrentovich]{Kim}
Y.-K. Kim, S.~V. Shiyanovskii and O.~D. Lavrentovich, \emph{J. Phys. Condens.
  Matter}, 2013, \textbf{25}, 404202\relax
\mciteBstWouldAddEndPuncttrue
\mciteSetBstMidEndSepPunct{\mcitedefaultmidpunct}
{\mcitedefaultendpunct}{\mcitedefaultseppunct}\relax
\EndOfBibitem
\bibitem[Abera \emph{et~al.}(2014)Abera, Verboven, Defraeye, Fanta, Hertog,
  Carmeliet, and Nicolai]{plant:2014}
M.~K. Abera, P.~Verboven, T.~Defraeye, S.~W. Fanta, M.~L. A. T.~M. Hertog,
  J.~Carmeliet and B.~M. Nicolai, \emph{Annals of Botany}, 2014, \textbf{114},
  605--617\relax
\mciteBstWouldAddEndPuncttrue
\mciteSetBstMidEndSepPunct{\mcitedefaultmidpunct}
{\mcitedefaultendpunct}{\mcitedefaultseppunct}\relax
\EndOfBibitem
\bibitem[Pollard(2004)]{Pollard}
T.~D. Pollard, \emph{J Exp Zool A Comp Exp Biol.}, 2004, \textbf{301},
  9--14\relax
\mciteBstWouldAddEndPuncttrue
\mciteSetBstMidEndSepPunct{\mcitedefaultmidpunct}
{\mcitedefaultendpunct}{\mcitedefaultseppunct}\relax
\EndOfBibitem
\bibitem[Eggert \emph{et~al.}(2006)Eggert, Mitchison, and Field]{Eggert}
U.~S. Eggert, T.~J. Mitchison and C.~M. Field, \emph{Annu. Rev. Biochem.},
  2006, \textbf{75}, 543--66\relax
\mciteBstWouldAddEndPuncttrue
\mciteSetBstMidEndSepPunct{\mcitedefaultmidpunct}
{\mcitedefaultendpunct}{\mcitedefaultseppunct}\relax
\EndOfBibitem
\bibitem[Green \emph{et~al.}(2012)Green, Paluch, and Oegema]{Green}
R.~A. Green, E.~Paluch and K.~Oegema, \emph{Annu. Rev. Cell Dev. Biol.}, 2012,
  \textbf{28}, 29--58\relax
\mciteBstWouldAddEndPuncttrue
\mciteSetBstMidEndSepPunct{\mcitedefaultmidpunct}
{\mcitedefaultendpunct}{\mcitedefaultseppunct}\relax
\EndOfBibitem
\bibitem[Fededa and Gerlich(2012)]{Fededa}
J.~P. Fededa and D.~W. Gerlich, \emph{Nat. Cell Biol.}, 2012, \textbf{14},
  29--58\relax
\mciteBstWouldAddEndPuncttrue
\mciteSetBstMidEndSepPunct{\mcitedefaultmidpunct}
{\mcitedefaultendpunct}{\mcitedefaultseppunct}\relax
\EndOfBibitem
\bibitem[Khodjakov and Rieder(2001)]{Khodjakov}
A.~Khodjakov and C.~L. Rieder, \emph{JCB Report}, 2001, \textbf{153},
  237--242\relax
\mciteBstWouldAddEndPuncttrue
\mciteSetBstMidEndSepPunct{\mcitedefaultmidpunct}
{\mcitedefaultendpunct}{\mcitedefaultseppunct}\relax
\EndOfBibitem
\bibitem[Lifshitz and Landau(1987)]{Landau6}
E.~Lifshitz and L.~Landau, \emph{Course of Theoretical Physics, volume VI:
  Fluid Mechanics}, Butterworth-Heinemann, 1987\relax
\mciteBstWouldAddEndPuncttrue
\mciteSetBstMidEndSepPunct{\mcitedefaultmidpunct}
{\mcitedefaultendpunct}{\mcitedefaultseppunct}\relax
\EndOfBibitem
\bibitem[{Ben Amar} \emph{et~al.}(2003){Ben Amar}, Cummings, and
  Pomeau]{benamar}
M.~{Ben Amar}, L.~J. Cummings and Y.~Pomeau, \emph{Phys. Fluids}, 2003,
  \textbf{15}, 2949\relax
\mciteBstWouldAddEndPuncttrue
\mciteSetBstMidEndSepPunct{\mcitedefaultmidpunct}
{\mcitedefaultendpunct}{\mcitedefaultseppunct}\relax
\EndOfBibitem
\bibitem[\mbox{Ben Amar} \emph{et~al.}(2011)\mbox{Ben Amar}, Manyuhina, and
  Napoli]{kera}
M.~\mbox{Ben Amar}, O.~V. Manyuhina and G.~Napoli, \emph{Eur. Phys. J. Plus},
  2011, \textbf{126}, 19\relax
\mciteBstWouldAddEndPuncttrue
\mciteSetBstMidEndSepPunct{\mcitedefaultmidpunct}
{\mcitedefaultendpunct}{\mcitedefaultseppunct}\relax
\EndOfBibitem
\bibitem[Prinsen and {van der Schoot}(2003)]{schoot}
P.~Prinsen and P.~{van der Schoot}, \emph{Phys. Rev. E}, 2003, \textbf{68},
  021701\relax
\mciteBstWouldAddEndPuncttrue
\mciteSetBstMidEndSepPunct{\mcitedefaultmidpunct}
{\mcitedefaultendpunct}{\mcitedefaultseppunct}\relax
\EndOfBibitem
\bibitem[Wincure and Rey(2007)]{rey}
B.~Wincure and A.~D. Rey, \emph{Continuum Mech. Thermodyn.}, 2007, \textbf{19},
  37\relax
\mciteBstWouldAddEndPuncttrue
\mciteSetBstMidEndSepPunct{\mcitedefaultmidpunct}
{\mcitedefaultendpunct}{\mcitedefaultseppunct}\relax
\EndOfBibitem
\bibitem[DoCarmo(1976)]{doCarmo}
M.~P. DoCarmo, \emph{Differential Geometry of Curves and Surfaces},
  Prentice--Hall, Englewood Cliffs, N.J., 1976\relax
\mciteBstWouldAddEndPuncttrue
\mciteSetBstMidEndSepPunct{\mcitedefaultmidpunct}
{\mcitedefaultendpunct}{\mcitedefaultseppunct}\relax
\EndOfBibitem
\bibitem[Manyuhina(2014)]{protein}
O.~V. Manyuhina, \emph{Phys. Rev. E}, 2014, \textbf{90}, 022713\relax
\mciteBstWouldAddEndPuncttrue
\mciteSetBstMidEndSepPunct{\mcitedefaultmidpunct}
{\mcitedefaultendpunct}{\mcitedefaultseppunct}\relax
\EndOfBibitem
\end{mcitethebibliography}
\bibliographystyle{rsc} 

\end{document}